\documentclass[acmsmall,screen]{acmart}

\AtBeginDocument{%
  }

\setcopyright{acmlicensed}
\copyrightyear{2018}
\acmYear{2018}
\acmDOI{XXXXXXX.XXXXXXX}
\acmConference[Conference acronym 'XX]{Make sure to enter the correct
  conference title from your rights confirmation email}{June 03--05,
  2018}{Woodstock, NY}
\acmISBN{978-1-4503-XXXX-X/2018/06}
\usepackage{graphicx}
\usepackage{algorithm}
\usepackage{algpseudocode}
\usepackage{wrapfig}
\usepackage{multirow}

\usepackage{tcolorbox}
\usepackage{xcolor}
\usepackage{colortbl}
\definecolor{ForestGreen}{cmyk}{0.91,0,0.88,0.12}
\usepackage{enumitem}

\usepackage{xspace}
\usepackage{soul}

\usepackage{pifont}
\usepackage{booktabs}

\usepackage{cleveref}
\usepackage{siunitx}
\usepackage{multirow}

\newif\ifshownotes
\shownotestrue

\newcommand{\SweBV}{SWE-bench-Verified\xspace}
\newcommand{\SweBP}{SWE-bench-Pro\xspace}
\newcommand{\SweBL}{SWE-bench-Lite\xspace}
\newcommand{\miniSA}{mini-SWE-Agent\xspace}
\newcommand{\icat}{iCAT-Agent\xspace}
\newcommand{\sonar}{Sonar Foundation Agent\xspace}
\newcommand{\oh}{OpenHands\xspace}
\newcommand{\claude}{Claude\xspace}
\newcommand{\gpt}{GPT\xspace}
\newcommand{\minimax}{MiniMax\xspace}
\newcommand{\exper}{ExpeRepair\xspace}
\newcommand{\cc}{Claude Code\xspace}
\newcommand{\haiku}{Claude Haiku\xspace}
\newcommand{\name}{DeskCheck\xspace}
\newcommand{\doc}{SWE-Doctor\xspace}

\newcommand{\cmark}{\textcolor{green!60!black}{\ding{51}}} % ✓
\newcommand{\xmark}{\textcolor{red}{\ding{55}}}            % ✗

\tcbset{
  examplebox/.style={
    colback=gray!5,
    colframe=gray!60,
    boxrule=0.5pt,
    arc=2pt,
    left=6pt,
    right=6pt,
    top=5pt,
    bottom=5pt,
    fontupper=\small
  }
}

\newtcolorbox{rqsummary}{
  examplebox,
  before skip=6pt,
  after skip=6pt
}

\begin{document}

%\title{Code Reasoning as a First-Class Citizen in Programming Agents}
\title{Enforcing Code Reasoning in Agentic Scaffold for More Effective Issue Resolution}

%%
%% Authors. The "anonymous" class option replaces this block with
%% "Anonymous Author(s)" in the submitted PDF, so the real details can stay
%% here -- but double-check the compiled PDF before submitting.
%% TODO: fill in the real author list before the camera-ready deadline.
\author[Changshu Liu, Ji young Kim, Alireza Ghazanfari, Saanvi Nallu, Reyhaneh Jabbarvand]{Changshu Liu, Ji young Kim, Alireza Ghazanfari, Saanvi Nallu, Reyhaneh Jabbarvand}
%\authornote{}
%\email{}
%\orcid{}
%\affiliation{%
%  \institution{}
%  \city{}
%  \state{}
%  \country{}
%}

\renewcommand{\shortauthors}{Anonymous et al.}

\begin{abstract}
%Code reasoning is increasingly used in both evaluating Large Language Models (LLMs) and applying them to programming tasks. 
Reasoning is essential to existing ReAct-based agents. In programming tasks such as issue resolution, the agent should reason about both textual artifacts and code.
Despite agents being deployed at scale to automate many programming tasks, it remains unclear whether they perform any form of code reasoning in their thoughts, or to what extent it helps. To shed light on the impact of explicit code reasoning in programming agents, we focus on issue-resolution tasks from \SweBV, \SweBP, and SWE-bench-Lite, and present a large-scale empirical study on \num{7955} trajectories from six single/multi-agent scaffolds across four programming languages. The results show that code reasoning, in $12$ different forms, is already part of the problem-solving process in $66.95\%$ of trajectories, though at varying intensity. Agents involving code reasoning in their trajectories achieve higher resolution rates, and they reason more intensively as task complexity increases.
% \jiyoung{Could we remove "throughout" to align align with RQ1/RQ2? We classify trajectory as reasoning-trajectory if at least once reasoning is observed. "thoughout" may imply that rasoning is consistently present across the entire trajectory. How about "Trajectories involving code reasoning achieve higher resolution rates."}. 
%These findings position code reasoning as a first-class citizen in both the design and evaluation of programming agents.
While the empirical study positions code reasoning as a first-class citizen for programming agents, agents perform it largely \emph{unguided} and \emph{spontaneous}, resulting in the following symptoms in their solutions: they may (1) reproduce only the surface-level symptoms of an issue during localization,
(2) fail to validate near-hit patches whose behavior is hard to exercise with tests,
(3) overlook requirements specified in the issue description, and
(4) insufficiently validate patches, with tests that lack behavioral diversity.

To address these limitations, we propose \name, an agentic scaffold that systematically enforces code reasoning at each stage of repository-level issue resolution:
\name simulates execution paths to trace the root cause during localization, simulates the execution of candidate patches to refine near-misses, audits patches against the requirements in the issue description, and reasons about inputs and expected outputs for corner cases before generating validation tests.
Evaluation on \num{731} \SweBP instances shows that \name improves the resolution rate over two widely used baselines, \miniSA and \cc, by $6.29\%$ to $10.53\%$, incurring only \$0.02--\$0.28 (average = \$0.14) additional cost per instance. A process-centric analysis shows that, compared to the baselines, \name consistently achieves higher recall at method-level localization, repairs more issues that were not precisely localized, and captures missed corner cases in near-hit patches that can be easily fixed and submitted successfully. 
The exact McNemar test and repeated experiments confirm that the observed superiority is not random ($p<10^{-10}$ for all subject LLMs) and is unlikely to change due to inherent nondeterminism in agent behavior.
These detailed per-phase analyses confirm that the improved outcome is not accidental but grounded in the design. 
%Moreover, patches revised in the audit and validation stages are significantly more likely to be resolved, confirming the contribution of guided code reasoning.

\end{abstract}

%%
%% CCS concepts. TODO: regenerate at https://dl.acm.org/ccs and paste the
%% resulting CCSXML block and \ccsdesc lines here.
\begin{CCSXML}
<ccs2012>
   <concept>
       <concept_id>10011007.10011074</concept_id>
       <concept_desc>Software and its engineering~Software creation and management</concept_desc>
       <concept_significance>500</concept_significance>
       </concept>
 </ccs2012>
\end{CCSXML}

\ccsdesc[500]{Software and its engineering~Software creation and management}

%% TODO: replace with the keywords for this paper.
\keywords{Programming agents, code reasoning, large language models}

\maketitle

\section{Introduction}
Reasoning is key to generalizability of Large Language Models (LLMs)~\cite {chollet2019measure,huang2023towards}, allowing them to apply learned knowledge to unseen tasks. Assessing it helps ensure LLMs are not overfitting to specific workflows or training data~\cite{wu2024reasoning,mirzadeh2025gsmsymbolic,zhang2024careful,liang2025swebenchillusion}, ultimately leading to more robust and reliable decisions.
% \reyhan{please cite more code reasoning papers in the following sentence from our community, as this may piss them off if review}
Code reasoning~\cite{liu2026codemind,chen2024runtime,liu2025evaluating,gu2024cruxeval,xie2026core, roy2026codesense, li2026laboratory,gao2026core,beger2025coconut}, which requires understanding syntax as well as program execution, generalizes LLMs' knowledge to a broad range of programming tasks, such as test generation, issue resolution, code translation, and program verification. 
%debugging~\cite{ding2024semcoder}, bug localization and prediction~\cite{liu2025assessing}, and program correctness verification~\cite{bouras2026hoareprompt}.
Beyond single-turn LLMs, programming agents~\cite{yang2024sweagent,wang2025openhands} autonomously navigate repositories and invoke tools, guided by reasoning about the problem specification and tool-execution observations. Prior work~\cite{liu2026online,liu2026plan,podivilov2026agentlens,bouzenia2025understanding} has analyzed trajectories of programming agents, focusing on the action or tool usage at each step, overlooking the underlying reasoning process that leads to actions. Specifically, there is a dearth of research on the analysis and impact of code reasoning. 
%Thus, an important question remains open: \emph{what is the role of code reasoning in the problem-solving process of programming agents?}

To mitigate this gap, we performed a large-scale empirical study (\S\ref{sec:empirical-study}) on the reasoning of programming agents performing repository-level issue resolution, 
which has become the most dominant task to evaluate the overall coding capability of agentic systems~\cite{jimenez2024swe,openai2024swebenchverified,deng2025swebenchpro,zan2025multiswebench}.
Specifically, we investigated \emph{the prevalence of different forms of code reasoning in issue resolution trajectories} (\textbf{RQ1}, \S\ref{sec:rq-empirical-prevalence}), its \emph{correlation with the resolution outcome} (\textbf{RQ2}, \S\ref{sec:rq-empirical-correlation}), and \emph{the factors triggering code reasoning in agents for issue resolution} (\textbf{RQ3}, \S\ref{sec:rq-empirical-factors}).
We systematically analyzed \num{7955} trajectories from six single- and multi-agent scaffolds on issue-resolution tasks from \SweBP, \SweBV, and \SweBL in Python, Go, TypeScript, and JavaScript.

We observed code reasoning in $66.95\%$ of trajectories, though to varying degrees. Code reasoning strongly correlates with resolution outcome. Agents turn to code reasoning \emph{more frequently} as problems become \emph{harder}. Strikingly, despite the importance of code reasoning and its strong positive correlation with the final outcome, agents use it spontaneously rather than in a systematic, guided way. We illustrate the need for code reasoning in different stages of the issue resolution task (localization, patching, and validation) with the following examples from the study:
% \jiyoung{Could we slightly tone down "spontaneously" here? Base on RQ1, what we directly observe is that code reasoning still appears even when it is requested only once in the system prompt rather than explicitly enforced at every step. We could describe as "agents exhibit code reasoning even when it is only broadly prompted rather than systematically enforced throughout the problem solving process". If we have additional evidence supporting the characterization as "spontaneous", it might be helpful to mention it here}. 
%Therefore, \emph{code reasoning may be shallow or absent precisely in the phases/steps where it is actually needed.}
A manual analysis of trajectories 
%(by three authors, individually and blindly performed with $\alpha=0.66$ inter-rater agreement) 
reveals four recurring limitations in different phases of issue resolution (localization, patching, and validation), each of which could have been avoided with a systematic incorporation of forms of code reasoning:

\begin{wrapfigure}{r}{0.5\textwidth}
    \vspace{-20pt}
    \centering
    \includegraphics[width=0.5\textwidth]{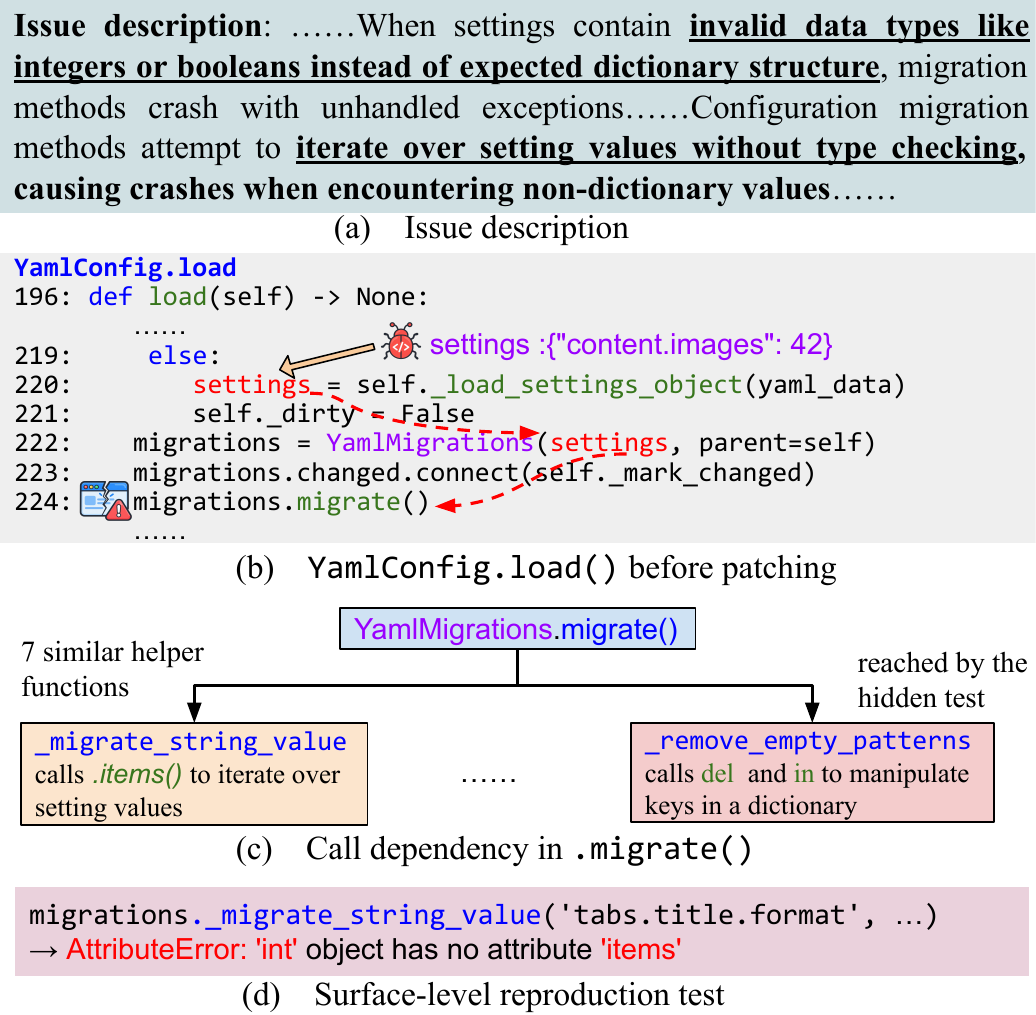}
    \vspace{-15pt}
    \caption{\cc(\haiku-4.5) failure to resolve \SweBP qutebrowser-8d05f028 due to surface-level issue reproduction.}
    \label{fig:reproduction-surface}
    \vspace{-30pt}
\end{wrapfigure}

\textbf{Limitation 1: Surface-level issue reproduction.}
Reproduction tests should turn issue descriptions into executable oracles for bug localization~\cite{yang2024sweagent,mundler2024swt,xia2024agentless}. However, many agents rely heavily on the issue description and generate tests that capture only surface-level symptoms, resulting in partial localization and incorrect patch. \Cref{fig:reproduction-surface}a shows an issue description from \SweBP, with the buggy code in \Cref{fig:reproduction-surface}b. The code initializes a \texttt{YamlMigrations} object from a \texttt{settings} in line 222, and calls the \texttt{migrate} function on this object in line 224. The \texttt{YamlMigrations} class consists of \emph{eight} helper methods (\Cref{fig:reproduction-surface}c), all of which are invoked by \texttt{YamlMigrations.migrate()}. The issue description says that when a \texttt{settings} object (and hence \texttt{migrations}) includes invalid data types such as a boolean or an integer instead of a dictionary, the code will crash. It does not explain which methods expect a dictionary type, but gives a hint about "iterating over settings values." \cc with \haiku-4.5 focuses on this overly specific hint, correctly localizing the bug to the \texttt{YamlMigrations} class, and writes reproduction unit tests for \emph{seven} out of \emph{eight} helpers, since they invoke \texttt{items()} API that iterates over \texttt{settings} values. Yet, it misses the \texttt{\_remove\_empty\_patterns} function, as its implementation does not involve iterations over the dictionary values, yet it still crashes on non-dictionary \texttt{settings}. 
%In \Cref{fig:reproduction-surface}a, \texttt{settings} is constructed with an integer value $42$ as the value in the dictionary (Line 220) and passed to \texttt{migrations} (Line 222), crashing inside \texttt{migrations.migrate()}, whose helpers all assume dictionary values (\Cref{fig:reproduction-surface}b). \cc focuses on the "iterate over setting values" in the issue description (\Cref{fig:reproduction-surface}c), reproduces only the \texttt{.items()} crash (\Cref{fig:reproduction-surface}d), and patches the seven helpers that call it, leaving \texttt{\_remove\_empty\_patterns} unchanged, which makes the same assumption via \texttt{in} and \texttt{del}. Although a reproduction test fails on the bug and passes on the patch, the patch does not survive the hidden tests.
%In fact, the root cause lies in \texttt{YamlConfig.load}, which constructs \texttt{settings} without guaranteeing the value types: a proper fix either checks the types before initializing \texttt{migrations} (Line 222) or validates them in \emph{all} helpers called in \texttt{migrations.migrate()}. 
\name successfully resolves this issue by reasoning about the call chains and execution paths involving (\texttt{YamlMigrations.migrate()}), localizing the root cause precisely and repairing the issue 

% \jiyoung{Could we align the description with the \ref{sec:qualitative}? There, \name traces the propogation of the invlaid setting through the execution path and identifies \texttt{\_remove\_empty\_patterns} as a previsouly missed edit site. How about "\name successfully resolves this issue by reasoning about the call chains and execution paths involving \texttt{YamlMigrations.migrate()}, tracing the propagation of the invalid setting and identifying previously missed edit site"?}.

%and considering In this example, the gap is  \textbf{the lack of reasoning from the symptom back to where it's produced.}

\begin{wrapfigure}{r}{0.51\textwidth}
    \vspace{-15pt}
    \centering
    \includegraphics[width=0.50\textwidth]{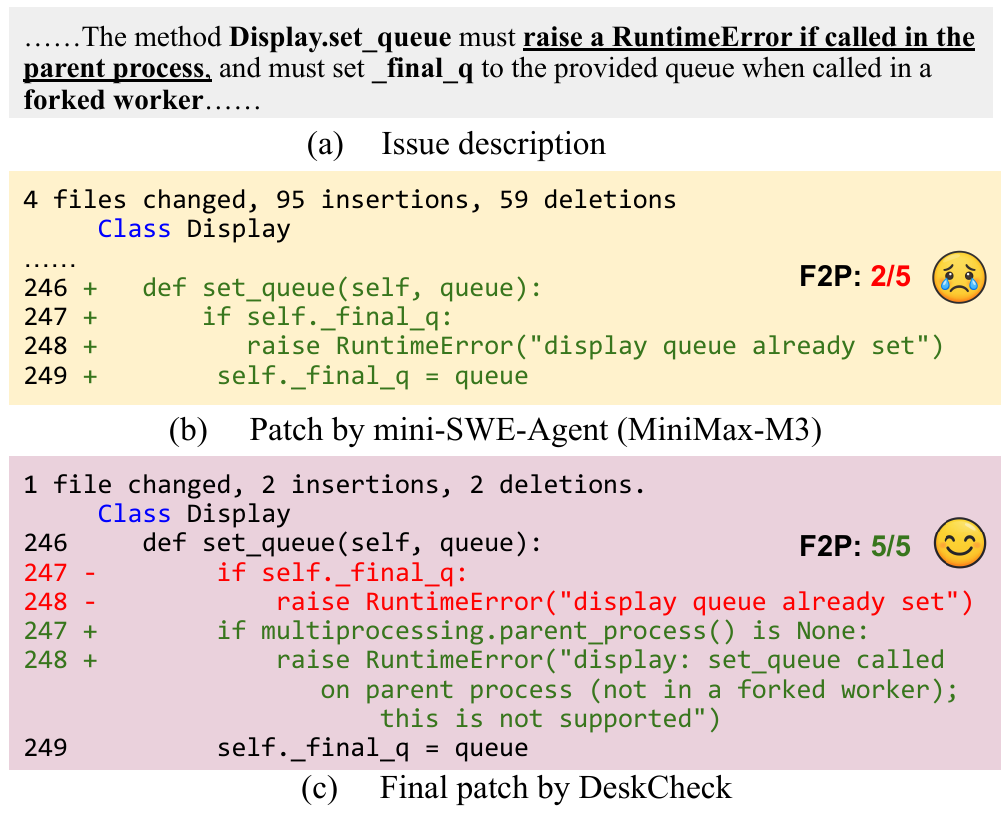}
    \vspace{-10pt}
    \caption{\miniSA failure to resolve \SweBP ansible-5e369604 due to difficulty of test generation.}
    \label{fig:near-miss}
    \vspace{-15pt}
\end{wrapfigure}

\textbf{Limitation 2: Test harness complexity.}
Covering the code for the purpose of validation or bug reproduction can be tricky, when reachability occurs under peculiar environmental settings or conditions. Such tests may require mocking or importing specific libraries solely for testing, not for the patch itself. \Cref{fig:near-miss} shows such a case, where per the issue description, the patch should implement a new function \texttt{set\_queue} in the \texttt{Display} class to set the class property \texttt{\_final\_q} to the provided queue. \emph{When this function runs in a parent process, it should terminate with a \texttt{RuntimeError}}. The \miniSA with \minimax-M3 implements the function as shown in \Cref{fig:near-miss}b, almost correctly but with an incorrect conditional check in line 247 (a near-hit patch~\cite{pabba2025refine,pmlr-v267-li25cf,liu2024automated, kim2026coherence, luo2026recap, wang2026solved}). It attempts to generate tests to run this function in a forked process for \emph{14 trajectory steps}, but fails to set up the environment properly to run the tests. It finally gives up and submits the patch without testing, passing only two of five golden tests. 
The correct fix requires replacing the previous condition with \texttt{multiprocessing.parent\_process() is None} (lines 247--248 in \Cref{fig:near-miss}c) to pass all five tests.
%\jiyoung{Could we keep this example at a slightly higher level? The \texttt{multiprocessing.parent\_process()} condition is introduced only here, it may be difficult to understand why is is the correct fix without reading the figure in detail. How about "A small refinement to this condition makes the patch pass all five test."?}
%\reyhan{@Ji young, thanks for the feedback. I would say more details are helpful here.}
%The gap here is \textbf{the absence of reasoning about how the produced patch itself behaves under the scenario the issue describes}.
Interestingly, \name generates the same initial patch, but refines it to the one in \Cref{fig:near-miss}c after simulating execution of the initial patch (without real execution; more details in \S \ref{sec:approach-repair-refine}) and realizing that the \texttt{RuntimeError} is not raised under the required condition. 
% \reyhan{@changshu, please update the reference}

%\textbf{\name successfully resolves this issue} () by simulating the execution of the initial patch, identifying that the expected error was not raised, and refining the error-handling logic accordingly. \reyhan{it is unclear why the agent was not able to produce the correct patch and why code execution simulation is helpful.}

\begin{wrapfigure}{r}{0.5\textwidth}
    \vspace{-10pt}
    \centering
    \includegraphics[width=0.5\textwidth]{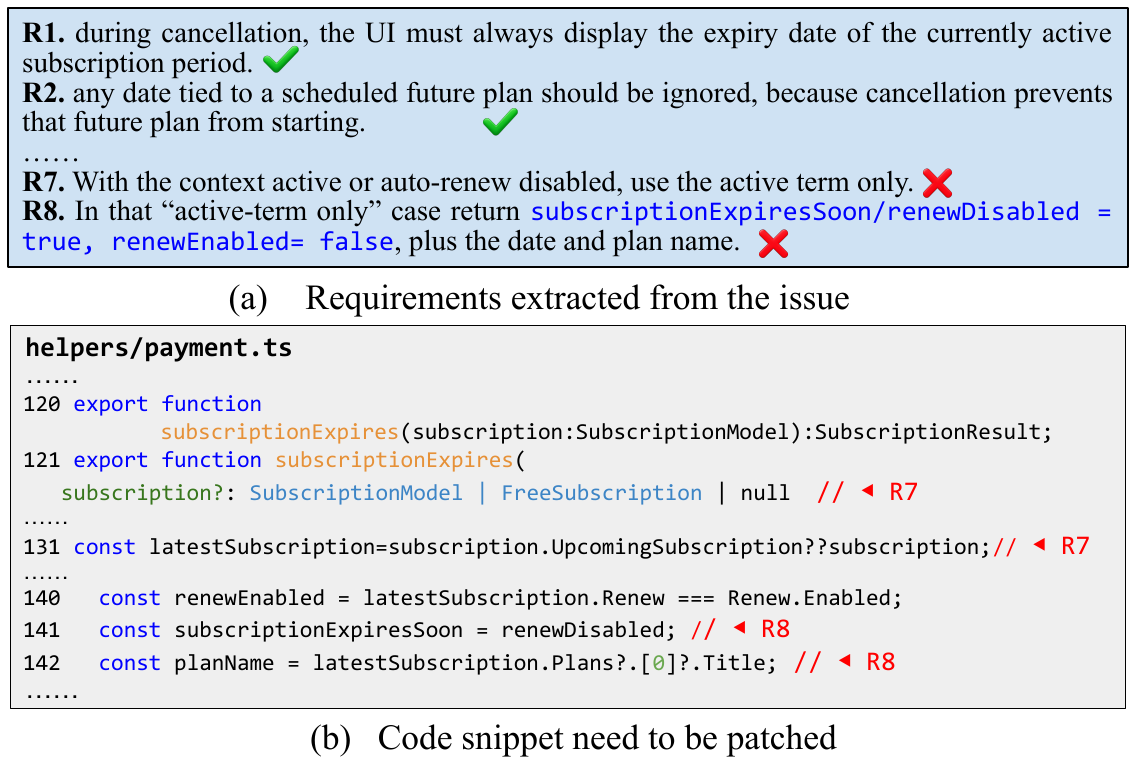}
    \vspace{-20pt}
    \caption{The buggy code corresponding to the specified requirements extracted from the issue description of \SweBP protonmail-ac23d1ef.}
    \label{fig:omitting-requirements}
    \vspace{-10pt}
\end{wrapfigure}
\textbf{Limitation 3: Neglecting requirements in the issue description.}
%make it relevant to specification reasoning: we assume that agents are understanding all aspects, but when the issue description is large with a lot of requirements, without specific focus on determining all the requirements, the patch may not reflect that. 
% the patch for this problem needs to change x files and y methods, and agent only changes x' and y'
% the issue is not like limitation 1 (where the agent focuses on a specific property metioned in the description) and not limitation 2 (where it understands the issues specification and even attempts to generate tests for it, but ignoring some requirements. We overcome this by making specification reasoning explicit through autdit and checks the patch against them
Issue descriptions can be long with multiple intertwined requirements. Although agents are instructed to understand the issue description, several studies show they overlook critical information in long issue descriptions~\cite{kuang2026reagent, zhou2026swerpg}. Therefore, they may generate incomplete patches that satisfy only a subset of issue description requirements. For example, \miniSA (\gpt-5.6-Luna), when attempting to resolve the problem protonmail-ac23d1ef from \SweBP, generates a partial patch (editing 15 out of 23 edit locations from the golden patch),
% \jiyoung{Could we clarify where the "23 edit locations" come from? I assme its from gold patch, then we could say "editing 15 out of 23 edit locations of the gold patch"}
satisfying some of the requirements in the long issue description~\cite{simagent2026artifact-Issue-Example}. 
% \reyhan{@changshu, please add a link to the issue description of this problem.}
With more explicit \emph{specification reasoning}~\cite{liu2026codemind}, \name prepares a list of \emph{eight} requirements from the issue description (\Cref{fig:omitting-requirements}a), and it simulates the execution of the generated patch under all requirements to confirm the patch satisfies all the requirements, resolving the issue. 
%\changshu{@Reyhan What actually happens with our approach is: \name prepares a list of \emph{eight} requirements from the issue description (\Cref{fig:omitting-requirements}a), and it simulates the execution of the generated patch under all requirements and finds the patch violates R7 and R8, and finally guides the agent to refine the patch.} \reyhan{this is something you explain later not here}

%Unlike near-hits, omitted requirements may leave entire edit sites untouched.
%\Cref{fig:omitting-requirements}a illustrates an issue description with eight requirements (R1--R8).
%\jiyoung{The issue contains eight requirements (R1--R8).} 
%The coding agent ships a patch satisfying \jiyoung{only} \num{6} out of \num{8} requirements(R1--R6) extracted in the issue description (\Cref{fig:omitting-requirements}a).
%The two omitted requirements R7 and R8 require non-trivial changes in lines \num{121} to \num{142} in \texttt{helpers/payment.ts}(\Cref{fig:omitting-requirements}b), which is never touched by the coding agent.
%The gap here is \textbf{the absence of requirement-by-requirement reasoning about how the completed patch behaves under each scenario the issue describes}, otherwise the omitted functionality goes unnoticed.
%\name resolves this issue by simulating the patch’s execution on inputs specified by $R7$ and $R8$, identifying the missing functionality, and refining the patch accordingly.

\begin{wrapfigure}{r}{0.5\textwidth}
    %\vspace{-5pt}
    \centering
    \vspace{-10pt}
    \includegraphics[width=0.5\textwidth]{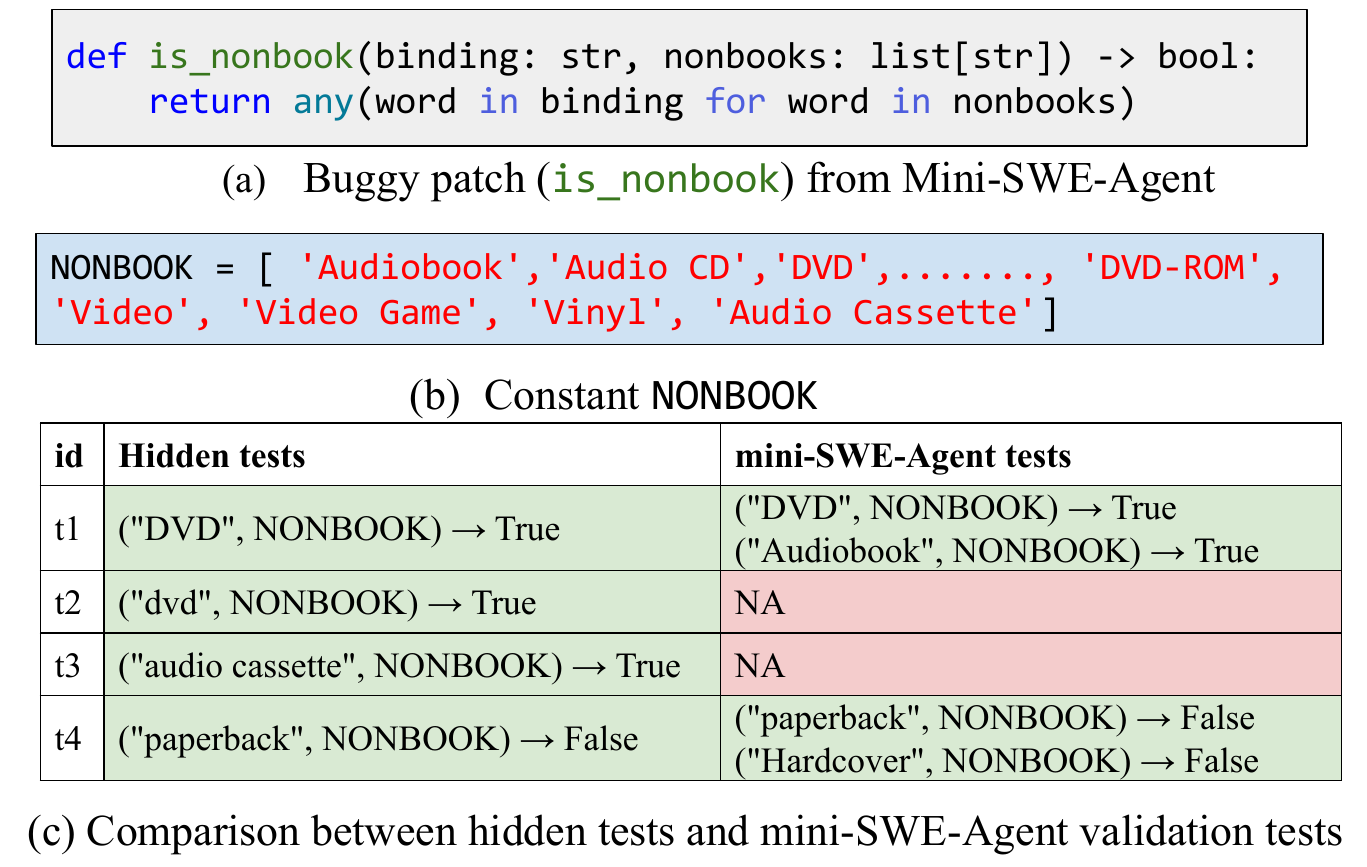}
    \vspace{-20pt}
    \caption{Comparison between SWE-bench-Pro hidden tests and
mini-SWE-Agent validation tests for issue openlibrary-3f7db6bb from \miniSA (\minimax-M3)}
    \label{fig:insufficient-validation}
    \vspace{-10pt}
\end{wrapfigure}

\textbf{Limitation 4: Insufficient validation.}
Issue resolution is complex, involving multiple sub-tasks: understanding the issue description, generating reproduction tests, patching, and generating validation tests for the new code (different from reproduction tests)~\cite{chen2026unlocking}. As a result, agents may not focus deeply on a single sub-task. Specifically, for test generation, agents usually reason about general corner cases of the newly generated code rather than maximizing behavioral diversity and covering the full input space.
% \jiyoung{Do we literally mean "covering the full input space", could we soften it to "broader set of behavioral conditions"?}.
\Cref{fig:insufficient-validation}a shows a patch generated by \miniSA (\minimax-M3) to resolve issue openlibrary-3f7db6bb from \SweBP. 
%Before submission, agents typically validate the patch with self-written tests~\cite{mundler2024swt,fei2026echo,guo2026swedoctor}, which makes the oracles encoded in those tests the determining factor for resolution: 
%An over-generous oracle lets the patch pass while it still fails the hidden tests~\cite{smith2015cure,wang2026solved}.
%\Cref{tab:limitations-oracles} \jiyoung{illustrates this problem for \texttt{\_is\_notebook} by} comparing the hidden-test oracles with those \miniSA generates on instance openlibrary-3f7db6bb.
The agent's generated tests (\Cref{fig:insufficient-validation}c) fall short not in number but in the variety of conditions they cover: they never consider case-insensitivity (t2) or multi-word (t3) inputs, resulting in failure. An enforced reasoning step to identify diverse inputs, determine their corresponding outputs helps \name resolve this issue. 
%The gap here is therefore \textbf{the absence of reasoning about test oracles beyond those the agent already had in previous phases}, without which the agent draws its confidence from oracles it invented to confirm itself. \jiyoung{The reference to "the oracles the agent already had in previous phase" may be little difficult to follow here. How about we emphasize the limited behavioral diversity of the generated test oracles?}
%\name broadens verification test coverage by reasoning about input–output pairs for possible corner cases before generating tests, yielding feedback that guides further patch refinement.

To overcome the mentioned limitations, this work proposes \name\footnote{Inspired by desk checking: programmers trace a program by hand with concrete inputs to find bugs without running it.}, an agentic scaffold that enforces various forms of code reasoning at different stages of issue resolution: in the \emph{Localization} phase, \name simulates the execution paths mined from the call graph to trace the root cause of the issue from surface-level symptoms in the issue description (\S\ref{sec:approach-localize});
%, without generating or running reproduction tests that require costly and failure-prone environment setup~\cite{bouzenia2025executionagent,hu2025repo2run,eliseeva2025envbench} (\S\ref{sec:approach-localize});
in the \emph{Repair \& Refine} phase, \name simulates the execution of the initial patch to check if it matches the expected behavior, and refines it based on the feedback collected from the simulation (\S\ref{sec:approach-repair-refine});
in the \emph{Audit} phase, \name extracts requirements from the issue description and leverages code reasoning to check if the generated patch fulfills all the requirements (\S\ref{sec:approach-patch-audit}); finally, 
in the \emph{Validation} phase, it reasons about a diverse set of input-expected output pairs to ensure a thorough validation (\S\ref{sec:approach-validation}).

Extensive evaluation of \name on  
%the \emph{effectiveness of \name} (\textbf{RQ4}, \S\ref{sec:effectiveness}), 
%\emph{the contribution of dedicated code reasoning in each stage} (\textbf{RQ5}, \S\ref{sec:effectiveness-stages}), 
%\emph{qualitative examples of its improvements} (\textbf{RQ6}, \S\ref{sec:qualitative}), 
%\emph{and an ablation on its design} (\textbf{RQ7}, \S\ref{sec:ablation}).
%We answer these RQs with extensive experiments on the 
\num{731} problems in \SweBP show consistent improvement in issue resolution compared to \miniSA ($+6.29\%$--$10.53\%$) and \cc ($+8.34\%$) baselines, agnostic to the studied backbone models, i.e., \minimax-M3, \gpt-5.6-Luna and \haiku-4.5. 
This comes with negligible overhead, increasing token usage by $7.6\%$ on average.
Every stage of guided code reasoning contributes to this gain (\textbf{RQ5}, \S\ref{sec:effectiveness-stages}).
Without executing reproduction tests, \name's localization achieves an average method-level recall of $0.66$, outperforming \miniSA ($0.46$), \cc ($0.38$), and \doc ($0.23$) with the same LLMs, even though \doc executes $28.0$ reproduction tests per instance on average.
Among the instances the baselines fail to resolve, patches revised in the Repair \& Refine, Audit, Validation phase by \name are more likely to be successfully resolved, which is supported by chi-square tests ($p<0.05$).
We also provide qualitative examples illustrating how guided code reasoning in each stage contributes to successful issue resolution (\S\ref{sec:qualitative}).
Finally, our ablation (\S\ref{sec:ablation}) shows that the resolution rate grows monotonically from $53\%$ to $68\%$ as code reasoning is incorporated stage by stage,
supporting that the improvement comes from code reasoning at different levels.

    Our notable contributions are: (1) a large-scale  study of 7,955 issue resolution trajectories (from six single- and multi-agent scaffolds resolving SWE-bench-family instances) to understand whether and how agents incorporate code reasoning in their problem-solving; 
(2) the design and implementation of \name, a code reasoning-driven multi-agent scaffold for issue resolution; and 
(3) extensive quantitative and qualitative evaluation of \name.

\section{Code Reasoning in Issue Resolution Agents}
\label{sec:empirical-study}
\begin{figure}[t]
    \vspace{-10pt}
    \centering
    \includegraphics[width=0.99\linewidth]{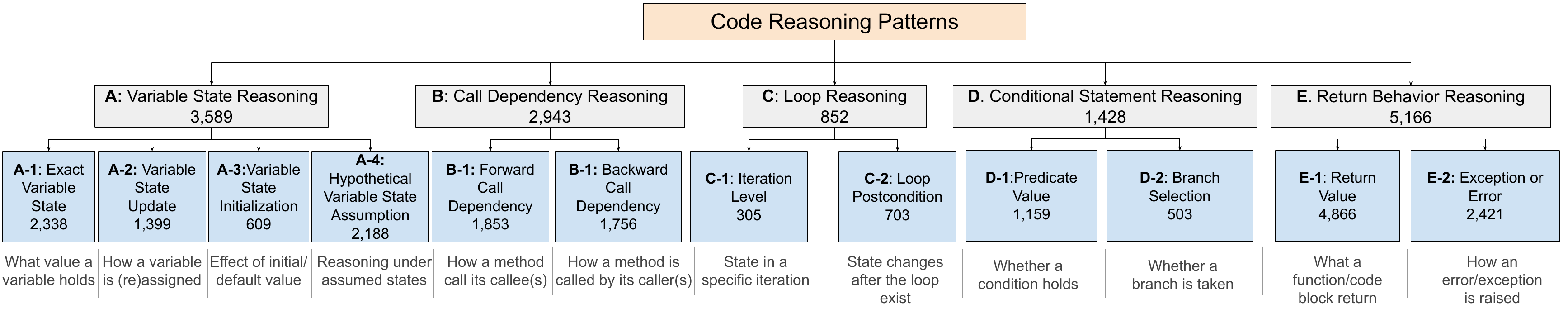}
    \vspace{-10pt}
    \caption{Code Reasoning Pattern Categorization.}
    \label{fig:categorization}
    \vspace{-10pt}
\end{figure}
%\subsection{Code Reasoning Pattern Taxonomy}

Our empirical study requires agent scaffolds that both publicly release trajectories on SWE-bench-family benchmarks and exhibit architectural diversity.
First, we anchor on two widely adopted scaffolds: \miniSA~\cite{yang2024sweagent} and \oh~\cite{wang2025openhands}, and complement them with scaffolds identified through a systematic search\footnote{The cut-off time is August 2026}  with the following criteria:
\emph{(I1):} Each trajectory records the agent's reasoning process, together with the action taken at each step, supporting process-centric analysis;
\emph{(I2):} We exclude model-training studies that reuse existing scaffolds; 
\emph{(I3):} We exclude scaffolds that share the core design of either anchored scaffold, including forks and minor variants; and
\emph{(I4):} The scaffold must either be introduced in a paper accepted to the main track of top AI/SE conferences in 2026 or 2025; released on arXiv in the past two months with a verifiable artifact; or have its results verified by an official leaderboard.

We locate \num{204} candidate scaffolds through keyword-matching\footnote{Keywords include "software engineering agent", "coding agent", "program repair", etc. We released the scripts for searching different sources in~\cite{simagent2026artifact-Search}.}, filter them based on the criteria above, and select the following four scaffolds: \exper~\cite{mu2026experepair}, \icat~\cite{chen2026unlocking}, \sonar~\cite{ruan2025sonar}, and \cc~\cite{chen2026unlocking}.
\exper is a program repair pipeline augmented with dual memory: an episodic memory of concrete past repair trajectories and a semantic memory of distilled insights, both consulted to guide repairs on new issues.
\icat is a multi-agent scaffold that achieves event-driven message-passing with a deliberately designed message bus. 
\sonar is a proprietary, closed-access coding-agent scaffold developed by Sonar, and \cc is a custom multi-agent system built with the Claude Code SDK released in \cite{chen2026unlocking}.
Across the six selected scaffolds, we analyze evaluations using backbone LLMs from two proprietary model families: \claude~\cite{anthropic2025claude} and \gpt~\cite{openai2026gpt5}, and one open-weight model family: \minimax~\cite{lai2026minimax}.

\subsection{Code Reasoning Patterns}

\textbf{Definition 1: Code Reasoning in Programming Agents:} an agent's reasoning about runtime behavior of a program fragment under a concrete or symbolic execution context, before taking any action to interact with the programming environment.
%Given the substantial human effort required to manually inspect natural-language reasoning traces, and the limited reliability of LLM-as-a-judge methods~\cite{wang2025trustjudge, bavaresco2025llms, stureborg2024large}, 
To automatically identify trajectory steps that appear to perform code reasoning, we build on prior work on evaluating code reasoning to identify code reasoning patterns~\cite{liu2026codemind,liu2025assessing, chen2024runtime, bouras2026hoareprompt}. 
% \reyhan{@changshu, please cite more papers from icse/fse related to code reasoning}
Commonly, the evaluation tasks in related work involve reasoning about variable state (A), call dependency (B), loop statements (C), conditional statements (D), and return statements/exceptions (E). First, we filter trajectory steps from the pool of selected trajectories whose reasoning\footnote{We analyze both the generated thoughts and the reasoning tokens, when an LLM enables access to them~\cite {anthropic_summarized_thinking}.} contains any code element, and prompt Claude Fable 5 to (1) determine whether that reasoning belongs to categories A--E and (2) assign it a more specific sub-category when possible, and (3) generate a regex pattern to search for similar patterns. The last step is especially important, as it lets us systematically search for those patterns again if the LLM missed them in individual assessments. \Cref{fig:categorization} presents the derived taxonomy and the number of trajectory steps whose reasoning matches the category/subcategory. Descriptions of each subcategory, with concrete examples, are available on the artifact website~\cite{simagent2026taxonomy}. 

\subsubsection{Manual Audit} To validate the accuracy of our LLM-based approach for reasoning categorization, we conduct a human audit on a randomly selected sample of \num{120} trajectories, comprising \num{60} classified as exhibiting code reasoning and \num{60} classified as not exhibiting it. 
The 120 trajectories were drawn at random in proportion to each agent–model combination, and they did not differ significantly from the full \num{7955} on any of nine characteristics, including benchmark, resolution status, task difficulty, and trajectory length\footnote{characteristics also include model, agent scaffold, code repository of the instance, number of steps detected with code reasoning, whether it contains reasoning tokens}, with $95\%$ confidence.
Three human labelers independently judged whether each trajectory involves code reasoning and, if so, which category it belongs to, reaching substantial agreement\footnote{measured with  Landis \& Koch scale~\cite{landis1977measurement}} (Krippendorff's $\alpha=0.66$~\cite{krippendorff1989content}).
Their \textbf{majority vote also substantially agrees with the automated approach ($\alpha=0.87$)}, confirming $59$ of the $60$ detected with code reasoning and $53$ of the $60$ detected without code reasoning.

\subsection{Empirical Results}
%We report the number of instances per code reasoning pattern in \Cref{fig:categorization}, e.g., \num{3589} for \emph{Variable State Reasoning}, \num{2943} for \emph{Call Dependency Reasoning}, and \num{4866} for \emph{Return Value}, the most prevalent (sub)category.
%We have released the empirical study artifact online~\cite{simagent2026artifactEmpirical}.

\begin{figure}[t]
    %\vspace{-10pt}
    \centering
    \includegraphics[width=0.95\linewidth]{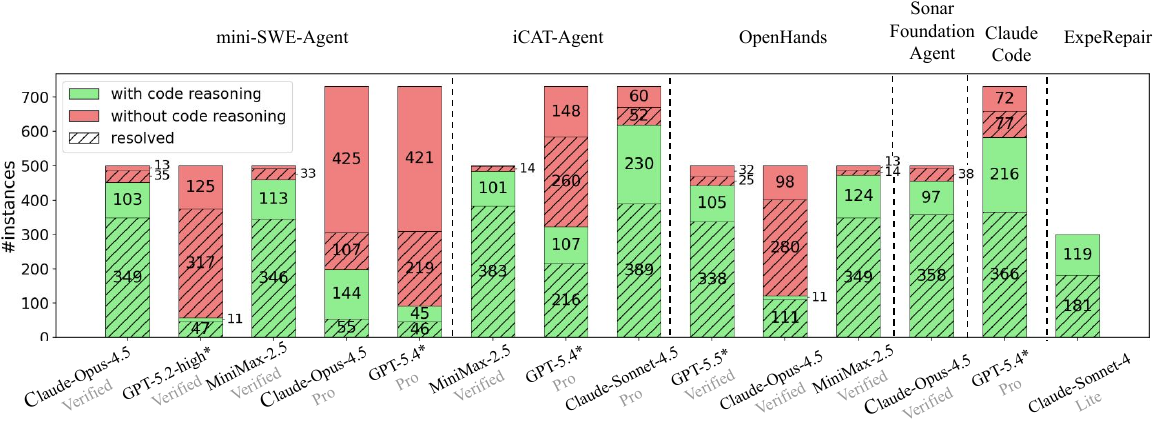}
    \vspace{-10pt} 
    \caption{Number of trajectories incorporating code reasoning. LLMs from OpenAI are denoted with * because their reasoning tokens are not publicly available. "Verified" indicates \SweBV, "Pro" indicates \SweBP, and "Lite" indicates \SweBL.}
    
    \label{fig:incorporating-code-reasoning}
    \vspace{-15pt}
\end{figure}

\begin{wraptable}[8]{r}{0.35\columnwidth}
\vspace{-20pt}
\centering
\footnotesize
\caption{Results of the chi-square test on the correlation between code reasoning and issue resolution per scaffold.}
\vspace{-10pt}
\label{tab:chi-test-empirical}
\setlength{\tabcolsep}{4pt}
\begin{tabular}{lcc}
\hline
\textbf{Scaffold} & \textbf{$\chi^2$} & \textbf{p-value} \\ \hline
\miniSA           & 183.0             & $1\times10^{-41}$ \\
\icat             & 5.31              & 0.021             \\
\oh               & 4.27              & 0.039             \\
\cc               & 3.91              & 0.048             \\ \hline
\end{tabular}
\vspace{-10pt}
\end{wraptable}

\subsubsection{\textbf{RQ1: Prevalence of Code Reasoning in Programming Agents.}}
\label{sec:rq-empirical-prevalence}
\Cref{fig:incorporating-code-reasoning} reports the number of trajectories where agents incorporate code reasoning while attempting to resolve issues.
% Across six agent scaffolds, the 14 agent configurations incorporate code reasoning in 66.95% of instances on average
Across six agent scaffolds, the \num{14} agent configurations incorporate code reasoning in $66.95\%$ of instances on average, regardless of whether the issues are ultimately resolved.
Different agents exhibit varying levels of code reasoning incorporation due to different scaffold designs.
\exper's prompts require an explicit analysis at every stage (e.g., fix analysis and patch review), so code reasoning appears in every trajectory, whereas \miniSA and \oh only request reasoning once in the system prompt, so it is less consistently incorporated due to agents' resistance to plan following~\cite{liu2026plan}.
These results suggest that \textbf{code reasoning can emerge during issue resolution even when it is not explicitly enforced at every stage of the agent workflow.}

\subsubsection{\textbf{RQ2: Correlation with resolution.}}
\label{sec:rq-empirical-correlation}
Across all the \num{14} sets of trajectories in \Cref{fig:incorporating-code-reasoning}, the resolution rate is $69.84\%$ when code reasoning is incorporated, compared to $50.81\%$ when it is not.
In \Cref{tab:chi-test-empirical}, a chi-square test~\cite{mchugh2013chi} over scaffolds with a sufficient number of instances (\sonar and \exper have too few samples without code reasoning detected) rejects the null hypothesis of no association between code reasoning and issue resolution, confirming that \textbf{code reasoning is associated with a higher resolution rate.}

\begin{figure}[H]
    \vspace{-12pt}
    \centering
    \includegraphics[width=0.99\linewidth]{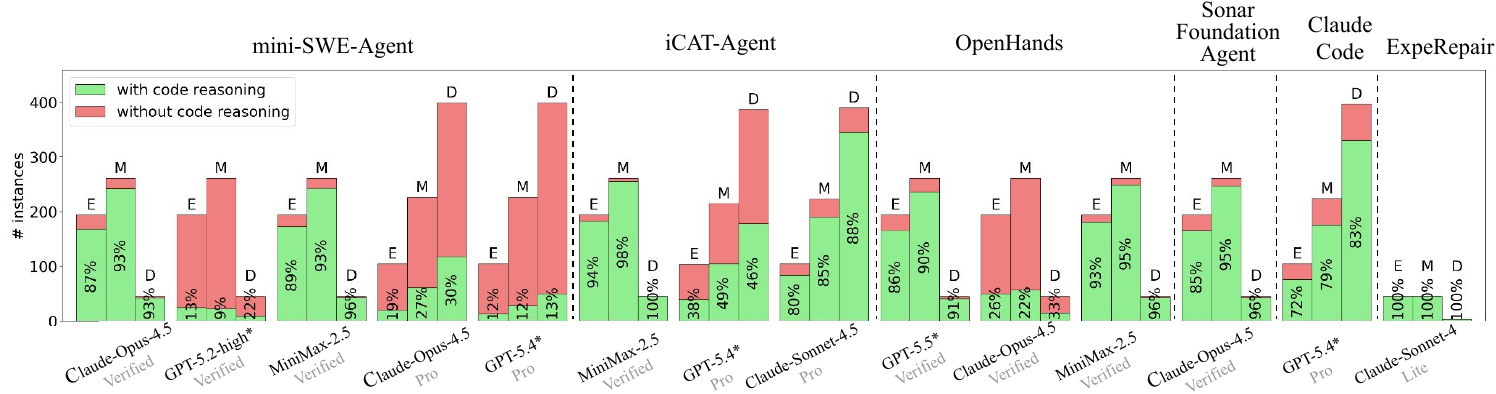}
    \Description{Chart showing the extent of code reasoning on different difficulty levels.}
    \vspace{-10pt}
    \caption{The extent of code reasoning on different difficulty levels.}
    \label{fig:difficulty-level}
    % \vspace{-10pt}
\end{figure}
\subsubsection{\textbf{RQ3: Influencing factors.}}
\label{sec:rq-empirical-factors}
\Cref{fig:difficulty-level} breaks down code reasoning by difficulty level, taken from the benchmark for \SweBV and \SweBL~\cite{jimenez2024swe} and determined by the number of edited files in the golden patch for \SweBP~\cite{zhang2026swe,chen2026unlocking}\footnote{Easy: 1 file; Medium: 2-3; Difficult: $\geq$ 4.}.
Averaged over the 14 agent configurations (equally weighted), code reasoning rises steadily from $63.9\%$ (easy) to $67.6\%$ (medium) and $70.5\%$ (difficult).
The trend is not strictly monotonic for every configuration: \miniSA(\gpt-5.2) and \oh(Claude-Opus-4.5) dip on medium instances ($13\% \rightarrow 9\% \rightarrow 22\%$ and $26\% \rightarrow 22\% \rightarrow 33\%$), while \icat(\gpt-5.4) peaks there ($38\% \rightarrow 49\% \rightarrow 46\%$).
% \jiyoung{I checked \Cref{fig:difficulty-level}, and the listed average values appear to be average across the 14 configurations (averaging the E,M, and D percentages separately). Text added to the sentence accordingly}.
% \changshu{I unified the way to report numbers in RQ1 and RQ3}
\textbf{This indicates that agents increasingly rely on reasoning about program behaviors when an issue becomes harder.}

% \subsubsection{Code Reasoning on Different Phases of Program Repair.}

\begin{figure}[H]
    \vspace{-10pt}
    \centering
    \includegraphics[width=0.99\linewidth]{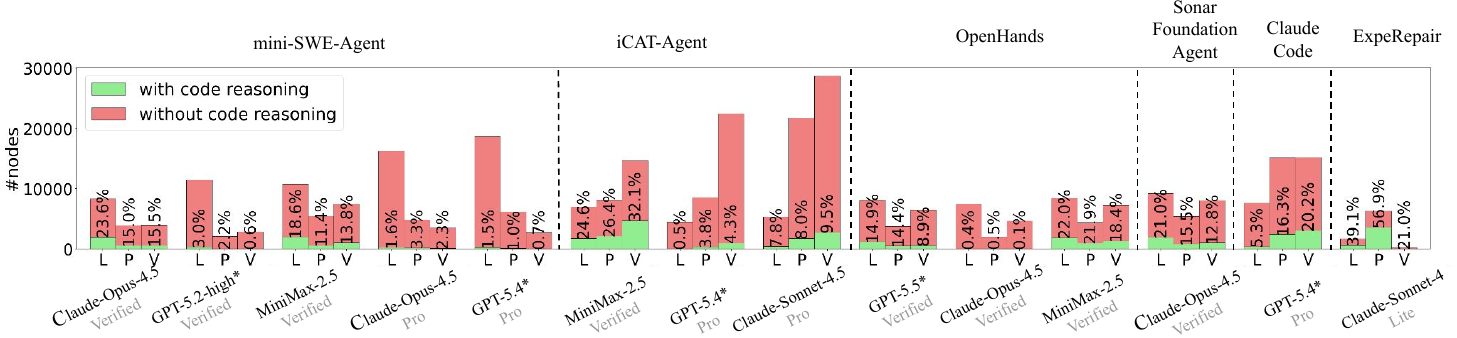}
    \Description{Chart showing the number of nodes incorporating code reasoning in the Localization phase (L), Patching phase (P), and Validation phase (V).}  % TODO: refine (ACM accessibility requirement)
    \caption{The number of nodes incorporating code reasoning in Localization phase (L), Patching  phase (P), and Validation phase(V). }
    \vspace{-5pt}
    \label{fig:phase}
\end{figure}

\Cref{fig:phase} shows the extent of code reasoning in each bug-fixing phase, where Graphectory~\cite{liu2026process} categorizes each action (node) in the trajectory into four phases: Localization (L), Patching (P), or Validation (V).
As in \Cref{fig:difficulty-level}, \textbf{scaffolds exhibit varying degrees of code reasoning across phases}: \miniSA, \oh, and \sonar reason most in Localization, \icat and \cc in Validation, and \exper in Patching.

Overall, code reasoning becomes more prevalent as issues become harder, yet when and where it occurs varies widely across scaffolds and phases.
\textbf{This spontaneous, uneven use of code reasoning motivates guiding it systematically at every phase of issue resolution, which is the design principle behind \name (\S\ref{sec:approach}).}

% \jiyoung{Could we add overall takeaway connecting \cref{fig:difficulty-level} and \Cref{fig:phase}? How about "Overall, these results suggest that code reasoning is not uniformly incorporated throughout issue resolution. While reasoning becomes more prevalent on average as task difficulty increases, this trend varies across agent configurations.". In addition to this, would it be helpful if we connect this observation back to the motivation in introduction that code reasoning currently occurs spontaneously rather than in as systematic guided manner by adding following sentence? "This variation further motivates systematically guiding code reasoning at different stages of the issue-resolution process."} \reyhan{These are great suggestions and I believe they can make the paper stronger}

% \begin{rqsummary}
% A comprehensive study across six scaffolds and three LLM families reveals that in $63.61\%$ of instances, coding agents naturally incorporate code reasoning in their bug-fixing processes. 
% This incorporation is associated with a significantly higher resolution rate ($69.31\%$) compared to those without code reasoning ($52.40\%$), and agents prefer to engage in code reasoning more often on harder issues.
% Meanwhile, the degree of code reasoning varies depending on the phase of the bug repair and the specific scaffold design.
% \end{rqsummary}

\begin{wrapfigure}{r}{0.52\linewidth}
\vspace{-10pt}
\setlength{\columnwidth}{\linewidth}%
\begin{algorithm}[H]
\algrenewcommand\textproc[1]{\textnormal{#1}}
\caption{Overall workflow of \name}
\label{alg:workflow}

\footnotesize
\begin{algorithmic}[1]
\Statex \textbf{Input:} Repository $\mathcal{R}$, issue description $I$
\Statex \textbf{Output:} Submitted patch $P$

\State $\mathit{canSubmit} \gets \textsc{False}$
\State $H \gets \Call{Explore}{\mathcal{R}, I}$
\State $\mathcal{E} \gets \Call{Localize}{\mathcal{R}, I, H}$
\State $P \gets \Call{RepairRefine}{\mathcal{R}, I, \mathcal{E}}$
\State $\mathit{hasViolation}, V \gets \Call{Audit}{\mathcal{R}, I, P}$
\If {$\mathit{hasViolation}$}
    \State $P \gets \Call{RepairRefine}{\mathcal{R}, I, V, \mathit{lightWeight}{=}\textsc{True}}$
\EndIf

\While {$\neg \mathit{canSubmit}$}
    \State $\mathit{err} \gets \Call{Validate}{\mathcal{R}, I, P}$
    \If {$\mathit{err} = 0$}
        \State $\mathit{canSubmit} \gets \textsc{True}$
    \Else
        \State $P \gets \Call{RepairRefine}{\mathcal{R}, I, \mathit{err}, \mathit{lightWeight}{=}\textsc{True}}$
    \EndIf
\EndWhile
\State \Return $P$

\end{algorithmic}
\end{algorithm}
\vspace{-35pt}
\end{wrapfigure}

\section{\name: Code Reasoning Augmented Issue Resolution Agent}
\label{sec:approach}

% Motivated by the prevalence of code reasoning in agent trajectories, even when scaffolds do not explicitly require it, and its positive impact on issue resolution, 
% we propose \name, which explicitly incorporates code reasoning into the problem-solving process of programming agents.

\textbf{Definition 2: Execution Simulation:} Given a program slice $s$ (e.g., an execution path or a patch), a concrete or symbolic input $i$, a programming agent $a$ reasons about the execution of $s$ step by step until it terminates, without executing it. Depending on $s$, a simulation may exhibit any subset of the \num{12} code reasoning patterns identified in \S\ref{sec:empirical-study}.
% \reyhan{looks good}

\begin{figure}[!t]
    \centering
    \includegraphics[width=0.90\linewidth]{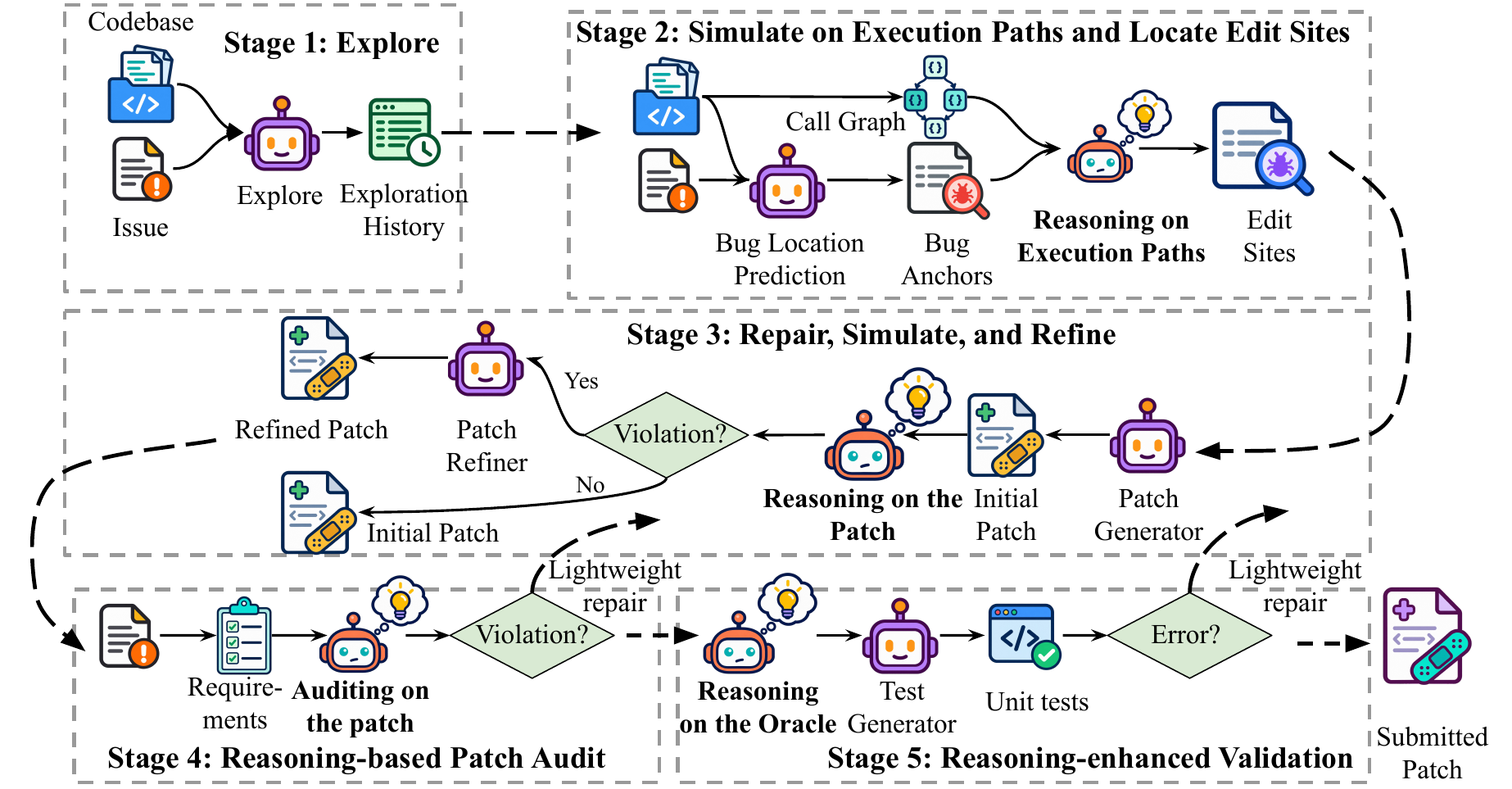}
    \vspace{-10pt}
    \caption{Overview of \name. Steps augmented with code reasoning are \textbf{bolded}.}
    \label{fig:approach_overview}
    \vspace{-15pt}
\end{figure}

\subsection{\name}
\label{sec:approach-design}
\Cref{fig:approach_overview} shows the workflow of \name, consisting of five stages: \emph{Explore}, \emph{Localize}, \emph{Repair \& Refine}, \emph{Audit}, and \emph{Validation}.
All stages except Explore are equipped with dedicated code reasoning components to improve their performance.
In \name, every stage is framed as a sub-agent to manage its context on specific tasks independently, alleviating possible context pollution~\cite{chen2026unlocking,liu2024lost,rajasekaran2025contextengineering}.
\Cref{alg:workflow} details the workflow: after exploring the repository (Line $2$), \name localizes the potential edit sites by simulating execution paths (Line $3$), generates and refines a patch $P$ (Line $4$) after simulating the patch, and uses code reasoning to audit $P$ against the requirements in the issue, routing any violations back to a lightweight repair (Lines $6$--$8$).
Finally, the Validation stage reasons about (input, expected output) pairs before generating and executing tests (Line $10$), and iteratively refines $P$ on test feedback until it decides to submit.

\subsubsection{Stage 1: Explore}
\label{sec:approach-explore}
\name starts with a read-only explore sub-agent that receives the full task specification, lists the project layout, greps for the symbols, error messages and API names it mentions, and reads the most likely involved source files, so that the next stage starts with the relevant code in hand.
It \emph{may not} edit, create or delete any file.

\begin{figure}[!t]
    \centering
    \includegraphics[width=0.99\linewidth]{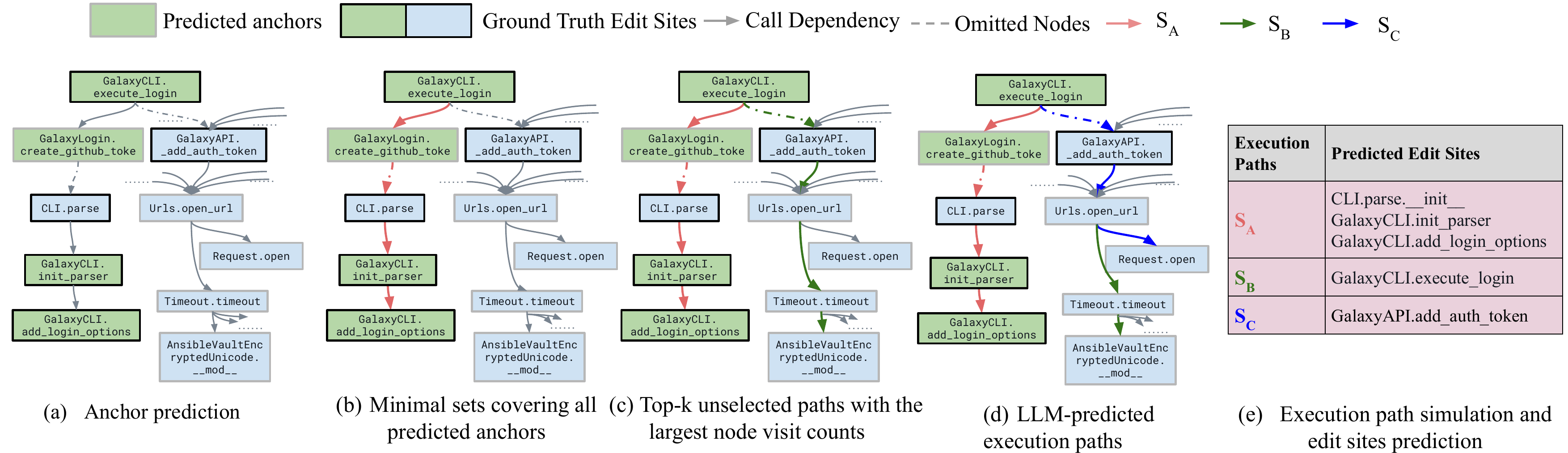}
    \caption{Example of the execution path selection and path simulation in the Localization stage.}
    \label{fig:localization}
    \vspace{-15pt}
\end{figure}

\begin{wrapfigure}{r}{0.45\linewidth}
\vspace{-15pt}
\setlength{\columnwidth}{\linewidth}%
\begin{algorithm}[H]
\algrenewcommand\textproc[1]{\textnormal{#1}}
\caption{Localization}
\label{alg:localization}
\footnotesize
\begin{algorithmic}[1]
\Statex \textbf{Input:} Repository $\mathcal{R}$, issue description $I$,
         LLM $M$, explore history $H$, expansion budget $k$
\Statex \textbf{Output:} Suggested edit sites $\mathcal{E}$

\State $G \gets \Call{BuildCallGraph}{\mathcal{R}}$
\State $\mathcal{P} \gets \Call{ConstructCandidatePaths}{G}$
\State $A \gets \Call{PredictAnchors}{M, I, \mathcal{R}, H, G}$

\State $\mathcal{S}_A \gets
    \Call{SelectAnchorCoveringPaths}{\mathcal{P}, A}$

\State $\mathcal{S}_B \gets
    \Call{TopK}{\mathcal{P} \setminus \mathcal{S}_A,
                k, p \mapsto |V(p)|}$

\State $\mathcal{S}_C \gets
    \Call{PredictBugRevealingPaths}{
        M, I, G}$
\State $\mathcal{S} \gets
    \mathcal{S}_A \cup \mathcal{S}_B \cup \mathcal{S}_C$

\State $\mathcal{E} \gets \emptyset$
\For {$s \in \mathcal{S}$}
    \State $Input \gets \Call{ConcreteInputPrediction}{M, I, s}$
    \State $e \gets \Call{SimulatePaths}{M, Input, s}$
    \State $\mathcal{E} \gets \mathcal{E} \cup e$
\EndFor
\State \Return $\mathcal{E}$

\end{algorithmic}
\end{algorithm}
\vspace{-25pt}
\end{wrapfigure}
\subsubsection{Stage 2: Localize}
\label{sec:approach-localize}
The Localization sub-agent takes the exploration history as input and identifies potential edit sites by reasoning on execution paths that may invoke the issue, \emph{without creating or running any reproduction test}, avoiding surface-level reproduction~\cite{fei2026echo} and costly environment setup~\cite{xiang2026evaluating}.
As shown in \Cref{alg:localization}, it first builds the call graph $G$ to capture the caller-callee relationships between modules and functions in the repository $\mathcal{R}$ and then extracts execution paths $\mathcal{P}$ from it (Lines $1$-$2$).
Then, given the exploration history and the issue description, \name predicts a set of anchors $A$, which may reveal the potential locations of the bug in the repository (Line $3$).
Subsequently, it selects a minimal set of paths $\mathcal{S}_A$ that cover \emph{all the predicted anchors} (Line $4$)\footnote{The set is built greedily, repeatedly adding the candidate path that covers the most still-uncovered anchors until all anchors are covered.}.
To ensure comprehensive coverage of potential bug locations, we expand it with the top-$k$ paths $\mathcal{S}_B$ with the largest node visit counts $V(p)$, i.e., the total number of times the nodes in path $p$ are visited by different paths (Line $5$), as nodes visited by many paths are more likely to lie on the execution flow leading to the bug.
After that, the LLM predicts the bug-revealing paths $\mathcal{S}_C$ based on the issue description, which are merged with $\mathcal{S}_A$ and $\mathcal{S}_B$ into $\mathcal{S}$ (Lines $6$-$7$).

With the selected paths, \name performs code reasoning on each execution path in $\mathcal{S}$: the LLM is instructed to (1) predict the inputs that would trigger the path, (2) simulate the execution step by step, and (3) flag the root cause and the edit sites if the simulation deviates from the expected behavior (Lines $9$-$12$), and returns the suggested edit sites $\mathcal{E}$ (Line $14$).
\name also supports new feature implementation and refactoring: since the expected behavior is derived from the issue description, simulating the base commit also captures missing functionality or structural changes, and the flagged edit sites mark where the requested behavior must be introduced.

% Algorithms 3 (Repair & Refine) and 4 (Patch Audit) side by side in one top float.
\begin{figure}[t]
\begin{minipage}[t]{0.48\linewidth}
\setlength{\columnwidth}{\linewidth}% algorithm[H] sizes itself to \columnwidth
\input{algotithm/repair-refine.tex}
\end{minipage}\hfill
\begin{minipage}[t]{0.48\linewidth}
\setlength{\columnwidth}{\linewidth}
\input{algotithm/audit.tex}
\end{minipage}
\vspace{-10pt}
\end{figure}

\Cref{fig:localization} illustrates a running example from \texttt{Ansible}.
The localization sub-agent first predicts four anchors covering three of the five ground truth edit sites (\Cref{fig:localization}a), and selects the red path covering all anchors into $\mathcal{S}_A$ (\Cref{fig:localization}b). 
Then it selects the green path with the largest node visit counts\footnote{We set k=1.} into $\mathcal{S}_B$ (\Cref{fig:localization}c), since \texttt{\_add\_auth\_token}, \texttt{open\_url}, and \texttt{timeout} are frequently visited nodes, and the blue bug-revealing path into $\mathcal{S}_C$ (\Cref{fig:localization}d).
The two previously missed edits in the anchor prediction are recovered step by step via the execution path selection, and 
after simulating these three paths, the LLM identifies all the ground truth edit sites (\Cref{fig:localization}e).

\begin{wrapfigure}{r}{0.48\linewidth}
\vspace{-25pt}
\setlength{\columnwidth}{\linewidth}% algorithm[H] sizes itself to \columnwidth
\input{algotithm/validate.tex}
\vspace{-20pt}
\end{wrapfigure}

\subsubsection{Stage 3: Repair \& Refine}
\label{sec:approach-repair-refine}
In practice, programming agents often produce patches that are partially correct but still contain \emph{flawed code logic}~\cite{pabba2025refine,pmlr-v267-li25cf,liu2024automated}.
To rescue such near-hits, the Repair \& Refine stage uses code reasoning to close the gap between a partially correct patch and a fully correct one (\Cref{alg:repair-refine}).
Given the suggested edit sites $\mathcal{E}$ from the Localization stage, the LLM $M$ first generates an initial patch $P$ (Line $4$), and then predicts a set of input-expected output pairs that fulfill both the implementation of $P$ and the issue requirements (Line $5$).
For each pair, $M$ simulates the execution of $P$ on the input and compares the resulting behavior against the expected output; on a mismatch, it reports a violation together with feedback \texttt{fb} describing the discrepancy (Lines $6$-$10$).
Finally, if any violation is found, an LLM refiner improves $P$ using the collected feedback $FB$ (Lines $13$-$14$).

\subsubsection{Stage 4: Patch Audit}
\label{sec:approach-patch-audit}
Another cause of failed patches is the issue description: agents may \emph{misinterpret or overlook critical requirements}, making no attempt to address them at all, especially in lengthy or ambiguous issue descriptions~\cite{zeng2025benchmarking,pabba2025refine}.
To address this, \name opens a clean context after Repair \& Refine,  and audits the patch against each requirement in the issue (\Cref{alg:patch-audit}).
It extracts the requirements with an LLM call (Line $4$), and for each requirement, it predicts an input-expected output pair, simulates the patch on the input, and collects feedback on any violation against the expected output (Lines $5$--$12$).
If any feedback is collected, \name performs a lightweight repair (Line $14$) that skips the full ``initial patch $\rightarrow$ simulate $\rightarrow$ refine'' flow to save tokens, directly fixing the patch based on feedback or refuting false alarms from the audit (\S\ref{sec:effectiveness-audit}).

\subsubsection{Stage 5: Validation}
\label{sec:approach-validation}
The validation tests may directly decide the outcome of an agentic pipeline: a high-quality validation test may disclose the defects in the generated patch and trigger further refinement if necessary.
To enrich the conditions covered by the validation tests, \name instructs the LLMs to propose possible corner cases and leverage code reasoning to generate input-expected output pairs for them and for reproducing the issue (Lines $3$--$4$ in \Cref{alg:patch-validate}).
These input-expected output pairs are turned into executable tests in Lines $7$-$10$, and executed with the regression tests if available in the repository.
If any test fails, \name triggers the lightweight patch refinement (Line $13$); otherwise, the validation sub-agent submits the final patch.

\vspace{-10pt}
\subsection{Experiment Setup}

\noindent\textbf{Implementation}.
We implement stages of \name on top of \miniSA and \cc, by modifying or injecting prompts into their original workflows, and set the localization expansion budget to $k=1$ to save tokens.
% The core prompts are available online ~\cite{simagent2026prompts}
Our prompts and implementation are publicly available~\cite{simagent2026prompts,simagent2026artifactpipeline}.

\noindent\textbf{Benchmark}.
We evaluate on \SweBP (\num{731} instances in Python, JavaScript, Go, and TypeScript), whose multi-file tasks are more challenging than those of \SweBV and \SweBL.
To prevent reward hacking, we disable network access and strip the git history in each Docker environment (the agent could otherwise recover the golden patch from the history).

\begin{wraptable}{r}{0.55\textwidth}
\vspace{-10pt}
\caption{Overall performance of \name compared to the baseline scaffolds on \SweBP.}
\vspace{-10pt}
\label{tab:overall_performance}
\centering
\scriptsize
\setlength{\tabcolsep}{2pt}
\renewcommand{\arraystretch}{0.95}
\begin{tabular}{cccc}
\hline
\textbf{Scaffold} & \textbf{Model} & \textbf{\%Resolved} & \textbf{Avg. Tokens/Cost} \\ \hline
\rowcolor[HTML]{EFEFEF}
\name   & \minimax-M3   & 65.25\% \textcolor{ForestGreen}{(+10.53)} & 5.61M/\$0.52 \\
\miniSA & \minimax-M3   & 54.72\%                                   & 5.27M/\$0.41\\
\rowcolor[HTML]{EFEFEF}
\name   & \gpt-5.6-Luna & 59.37\% \textcolor{ForestGreen}{(+6.29)}  & 1.42M/\$0.10 \\
\miniSA & \gpt-5.6-Luna & 53.08\%                                   & 1.65M/\$0.08 \\
\rowcolor[HTML]{EFEFEF}
\name   & \haiku-4.5    & 51.16\% \textcolor{ForestGreen}{(+8.34)}     & 6.75M/\$1.59      \\
\cc     & \haiku-4.5    & 42.82\%                                          &  5.89M/\$1.31     \\ \hline
\end{tabular}
\vspace{-12pt}
\end{wraptable}

\noindent\textbf{Subject LLMs and Baselines}.
We compare \name built on \miniSA against vanilla \miniSA with \minimax-M3 and \gpt-5.6-Luna.
Since \cc has no official \SweBP harness, we build a ``localize $\rightarrow$ repair $\rightarrow$ validate'' scaffold that runs the \cc API inside each instance's Docker container, and compare it with \name built on \cc using \haiku-4.5.
All LLMs use high reasoning effort.

\vspace{-5pt}
\subsection{\textbf{RQ4. Effectiveness of \name in Issue Resolution}}
\label{sec:effectiveness}
\subsubsection{Overall Performance of \name}

\begin{wrapfigure}{r}{0.5\columnwidth}
    \vspace{-5pt}
    \centering
    \includegraphics[width=\linewidth]{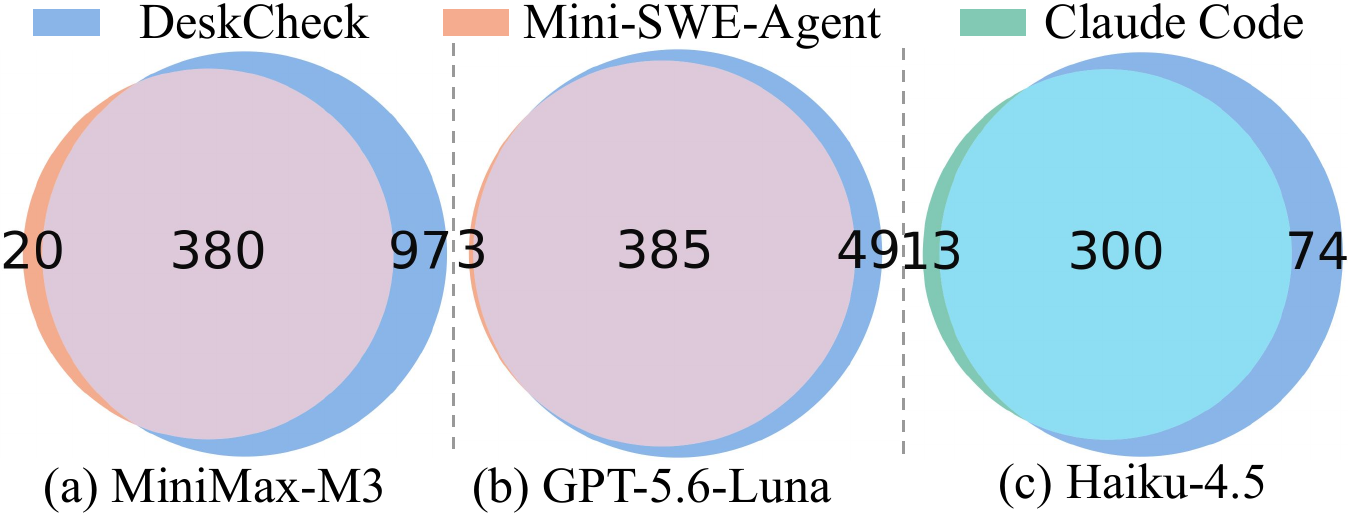}
    \caption{Comparison on unique fixes between \name, \miniSA, and \cc.}
    \label{fig:venn}
    \vspace{-15pt}
\end{wrapfigure}

\Cref{tab:overall_performance} reports the overall performance of \name on the \num{731} problems in \SweBP.
\name resolves $65.25\%$, $59.37\%$, and $51.16\%$ of instances with \minimax-M3, \gpt-5.6-Luna, and \haiku-4.5, respectively, \textbf{outperforming the \miniSA and \cc counterparts by $\mathbf{10.53\%}$, $\mathbf{6.29\%}$, and $\mathbf{8.34\%}$}.
We further visualize the unique fixes from different scaffolds in \Cref{fig:venn}.
Compared to \miniSA, \name generates $97$ unique fixes when loaded with \minimax-M3, and $49$ unique fixes when powered by \gpt-5.6-Luna.
Compared to \cc, \name generates $74$ unique fixes that are never produced by \cc.
\textbf{\name is cost-effective as well.} 
We report the average token consumption and cost in dollars per instance in \Cref{tab:overall_performance}.
\name consumes comparable amounts of tokens to \miniSA and \cc (at most $14.6\%$ more with \haiku-4.5),
 and costs only $\$0.02$--$\$0.28$ more per instance, in exchange for a notably higher resolution rate.
The replication package for this RQ is publicly available~\cite{simagent2026artifactOverall}.

\subsubsection{Breakdown by Difficulty Levels.}
\begin{wrapfigure}{r}{0.5\columnwidth}
    \vspace{-10pt}
    \centering
    \includegraphics[width=0.5\columnwidth]{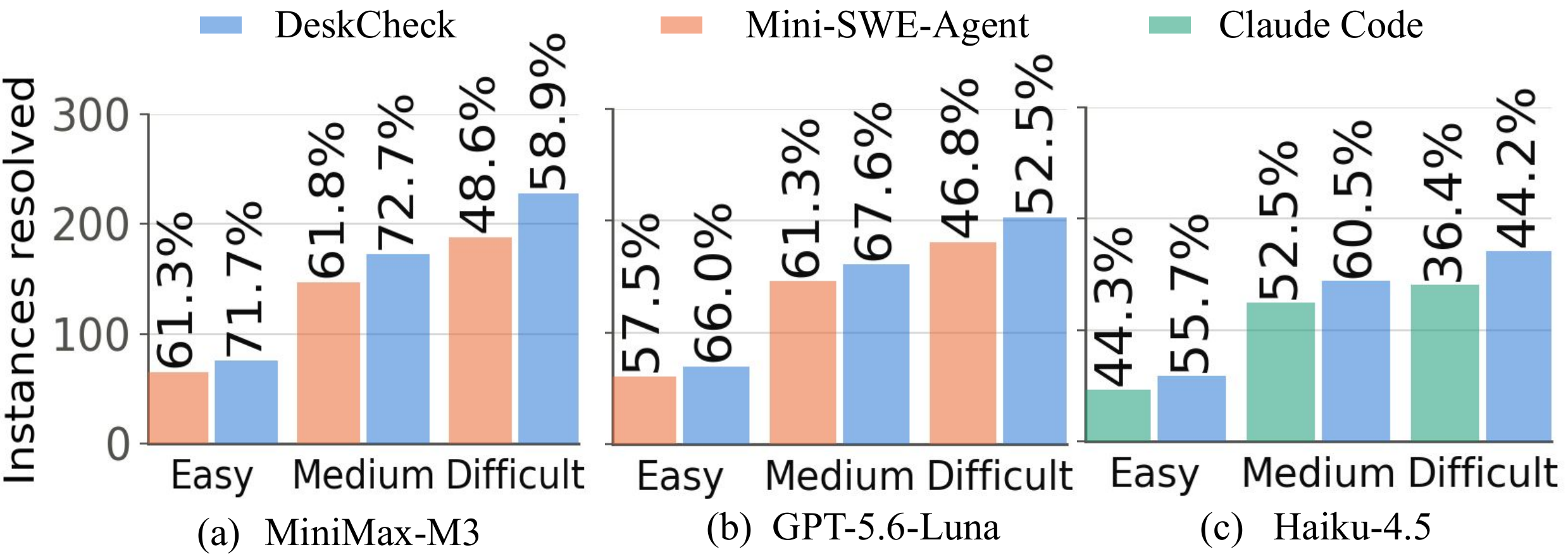}
    \vspace{-15pt}
    \caption{Difficulty breakdown of resolved instances.}
    \label{fig:difficulty-breakdown}
    \vspace{-15pt}
\end{wrapfigure}

As explained in \S\ref{sec:rq-empirical-factors}, we use the golden patch characteristics as a proxy for each problem's difficulty level.
\Cref{fig:difficulty-breakdown} shows the breakdown of resolved instances by difficulty level.
We observe that \name \textbf{consistently improves the resolution rate across all difficulty levels} compared to the baseline scaffolds.
Compared to \miniSA, on average \name resolves $9.4\%$, $8.7\%$, and $8.0\%$ more instances for easy, medium, and difficult problems, respectively.
\name resolves $11.4\%$, $8.0\%$, and $7.8\%$ more instances for easy, medium, and difficult problems compared to \cc.

\vspace{-5pt}
\subsubsection{Breakdown by Programming Languages}
\begin{figure}[!h]
\vspace{-10pt}
\centering
\includegraphics[width=0.90\linewidth]{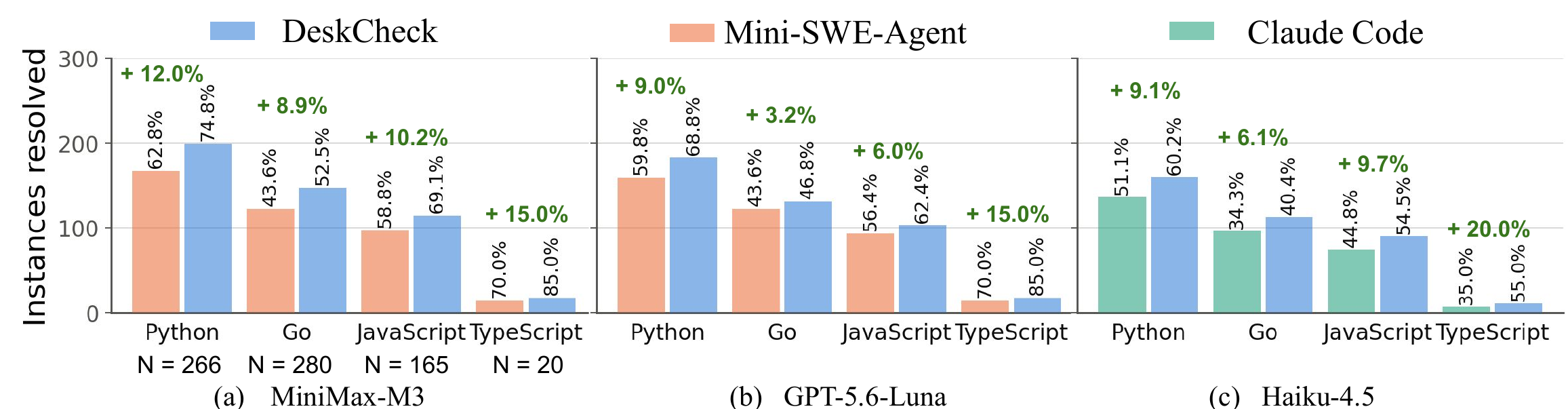}
\Description{Chart showing the breakdown of resolved instances by programming language.}
\vspace{-10pt}
\caption{Breakdown of resolved instances by programming language. }
\label{fig:pl-breakdown}
\vspace{-20pt}
\end{figure}

\Cref{fig:pl-breakdown} shows that \textbf{\name consistently improves the resolution rate across all programming languages} compared to the baseline scaffolds.
Go is the most challenging language for all three scaffolds and the one where \name gains the least; our human verification (\S\ref{sec:human-verification}) attributes this largely to \SweBP's grading tests, which more often require symbol names and method names that appear only in the golden patch but never mentioned in the issue description (e.g., \Cref{fig:fals-negative}).

\subsubsection{Impact of Non-determinism}
\label{sec:non-determinism}
To rule out LLM non-determinism, we apply the exact McNemar test~\cite{mcnemar1947note} on per-instance resolution status, whose null hypothesis is that \name and the baseline differ only by run-to-run randomness, i.e., each fixes roughly as many instances the other fails.
The test rejects it with \minimax-M3 ($p=2.6\times10^{-13}$), \gpt-5.6-Luna ($1.0\times10^{-11}$), and \haiku-4.5 ($1.6\times10^{-11}$), confirming that the improvements are statistically significant.
We further rerun the baselines twice more on the instances that only \name resolves (97, 49, and 74 for \minimax-M3, \gpt-5.6-Luna, and \haiku-4.5 from \Cref{fig:venn}), and they resolve only $16$, $7$, and $14$ of them in at least one retry.
Hence, $83.18\%$ of \textbf{these gains likely stem from \name itself rather than run-to-run non-determinism}.

% \begin{rqsummary}
% Finding 2: \name outperforms the baseline \miniSA and \cc by $8.41\%$ and \hl{XX}, respectively on \SweBP, while consuming similar amounts of tokens.
% Meanwhile, its improvement is consistent on different difficulty levels and programming languages.
% \end{rqsummary}

\subsubsection{Manual Analysis of Generated Patches}
\label{sec:human-verification}

\begin{wrapfigure}{r}{0.55\columnwidth}
    \vspace{-10pt}
    \centering
    \includegraphics[width=0.55\columnwidth]{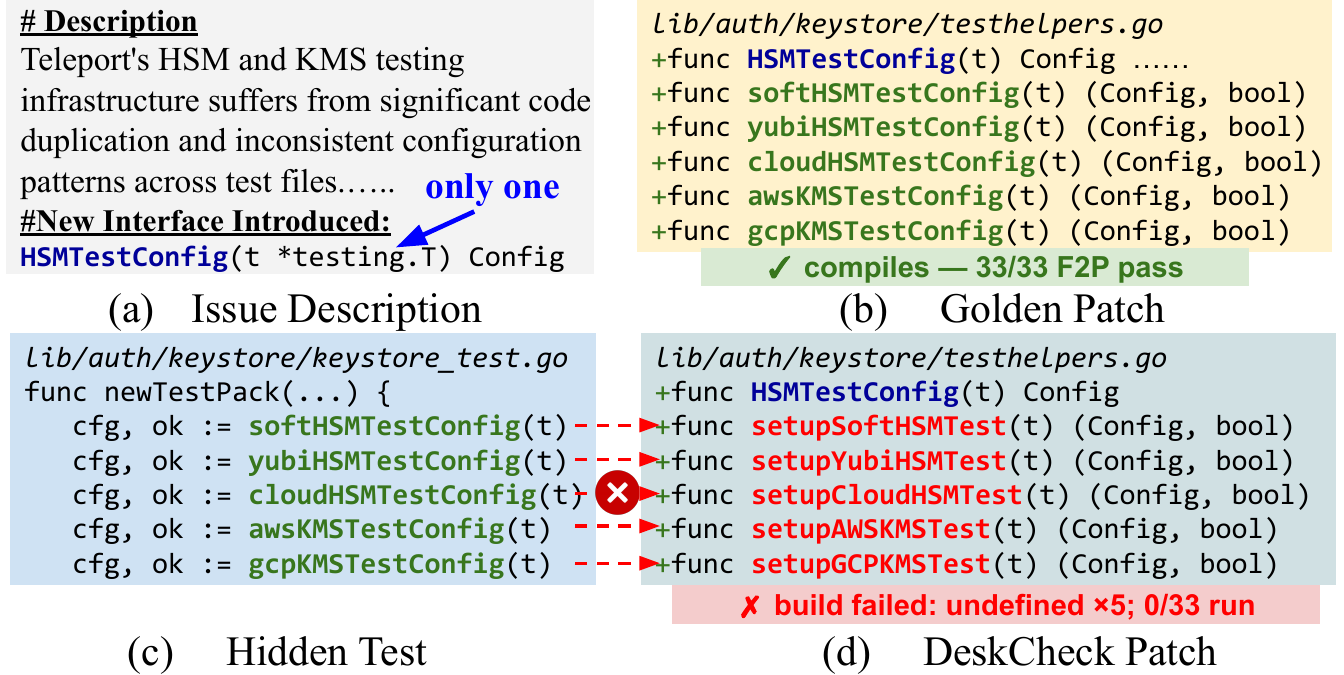}
    \vspace{-20pt}
    \caption{An example of hidden tests requiring function/symbol definitions only appear in the golden patch.}
    \label{fig:fals-negative}
    \vspace{-15pt}
\end{wrapfigure}

Failing the grading tests of \SweBP does not necessarily mean an issue is unresolved: SWE-bench-style tests can be overly specific and reject correct patches that deviate from the golden patch's implementation details~\cite{wang2026solved,yu2025utboost,aleithan2024swebenchplus,openai2024swebenchverified}.
We therefore have three human evaluators independently verify, against the issue description, the patches for the $234$ instances that both \name and \miniSA fail with \minimax-M3, reaching substantial agreement (Krippendorff's $\alpha=0.70$) and judging $94$ (\name) and $96$ (\miniSA) of them as correct.
Such false negatives arise when the hidden tests (1) contradict the issue description, (2) cover conditions never specified in the issue, (3) require function/symbol definitions that only appear in the golden patch, or (4) rely on fixtures/configurations that only exist at grading stage.
Of the $190$ false negatives, $32$, $125$, $31$, and $2$ are Python, Go, JavaScript, and TypeScript instances, respectively, which largely explains why both scaffolds struggle on Go (\Cref{fig:pl-breakdown}).

\Cref{fig:fals-negative} presents such a Go instance: the issue (\Cref{fig:fals-negative}a) names only one new function, \texttt{HSMTestConfig}, but the hidden test (\Cref{fig:fals-negative}c) calls five private helpers by their golden-patch names (\Cref{fig:fals-negative}b; \texttt{softHSMTestConfig}, \texttt{yubiHSMTestConfig}, etc.). 
\name implements the same design but names its helpers differently (\texttt{setupSoftHSMTest}, \texttt{setupYubiHSMTest}, etc.), so the test package fails to compile and none of the 33 hidden tests run. 
These non-trivial false negatives suggest that \textbf{resolution rates on \SweBP underestimate actual capability}, calling for caution when comparing approaches by small margins.
\subsection{\textbf{RQ5: Contribution of 
Guided Code Reasoning to Each Stage of \name}}
\label{sec:effectiveness-stages}
\subsubsection{Code Reasoning in the Localization}
To evaluate code reasoning in the Localization stage, we compare the method-level edit sites predicted by the Localization sub-agent with the methods/functions modified in the golden patch, which serve as a \emph{proxy} for the ground truth\footnote{This proxy may contain false positives: the golden patch can include edits beyond the issue, and a bug may admit multiple valid fixes.}.
We report both file- and method-level recall, precision, and F1 against the ground truth.

% \usepackage{multirow}
\begin{wraptable}{r}{0.5\textwidth}
\vspace{-5pt}
\footnotesize
\centering
\setlength{\tabcolsep}{3.5pt}
\caption{Localization Recall(Rec), Precision(Prec), and F1 of \name, \miniSA, \cc, and \doc on $107$ Python instances. \%Res refers to the resolution rate.}
\label{tab:localization}
\vspace{-5pt}
\resizebox{0.5\textwidth}{!}{%
\begin{tabular}{llc|ccc|ccc}
\hline
\multirow{2}{*}{\textbf{Agent}} & \multirow{2}{*}{\textbf{LLM}} & \multirow{2}{*}{\textbf{\%Res}} & \multicolumn{3}{c|}{\textbf{Method-Level}} & \multicolumn{3}{c}{\textbf{File-Level}} \\ \cline{4-9}
 & & & \textbf{Rec.} & \textbf{Prec.} & \textbf{F1} & \textbf{Rec.} & \textbf{Prec.} & \textbf{F1} \\ \hline
\name-L   & \multirow{5}{*}{\rotatebox[origin=c]{90}{MiniMax-M3}} & \textbf{73.83}  & \textbf{0.69} & 0.71 & \textbf{0.66} & 0.84 & 0.84          & 0.79         \\
\name-P        &                                & \textbf{73.83}   & 0.65   & 0.73   & 0.64   & \textbf{0.85}   & 0.88   & 0.82   \\
\miniSA-P &                                & 57.94   & 0.46          & 0.62          & 0.51          & 0.81          & \textbf{0.95} & \textbf{0.87} \\
\doc-L    &                                & 60.75   & 0.18          & \textbf{0.89}         & 0.30          & 0.49          & 0.73          & 0.53          \\
\doc-P        &                                & 60.75   &0.63    & 0.76   & \textbf{0.66}  & 0.71   & 0.88   & 0.76   \\ \hline
\name-L   & \multirow{5}{*}{\rotatebox[origin=c]{90}{GPT-5.6-Luna}} & \textbf{66.36}   & \textbf{0.67} & 0.66 & \textbf{0.63} & 0.88 & 0.87          & 0.84          \\
\name-P        &                                & \textbf{66.36}   & 0.64   & 0.69   & 0.62   & 0.89   & 0.88   & 0.85  \\
\miniSA-P &                                & 57.01  & 0.46          & 0.55          & 0.48          & \textbf{0.90}          & \textbf{0.97} & \textbf{0.91} \\
\doc-L    &                                & 58.88   & 0.28          & \textbf{0.81}          & 0.38          & 0.53          & 0.93          & 0.64          \\
\doc-P        &                                & 58.88   & 0.59   & 0.76   & 0.62   & 0.67   & 0.89   & 0.74   \\ \hline
\name-L   & \multirow{4}{*}{\rotatebox[origin=c]{90}{Haiku-4.5}} & \textbf{58.88}   & \textbf{0.61}              & 0.83              & \textbf{0.67}              &  \textbf{0.82}             &    \textbf{0.97}           &  \textbf{0.86}             \\
\name-P        &                                & \textbf{58.88}   & 0.60   & \textbf{0.84}   & 0.66   & \textbf{0.82}   & \textbf{0.97}   & \textbf{0.86}   \\
\cc-L     &                                & 46.72   & 0.38   & 0.54   & 0.41   & 0.81   & 0.94   & 0.84          \\
\cc-P        &                                & 46.72   & 0.58   & 0.80   & 0.63   & 0.81   & 0.96   & 0.85   \\ \hline
\end{tabular}%
}
\vspace{-10pt}
\end{wraptable}

We compare against three reproduction test-based approaches: \miniSA and \cc, both instructed to locate buggy sites with reproduction tests, and \doc~\cite{guo2026swedoctor}.
\doc executes multi-faceted reproduction tests, creates runtime-grounded diagnosis records (e.g., suspected fault location and observed runtime values), and feeds them to a downstream patch generator (\miniSA in its implementation), outperforming reproduction test generation approaches such as e-Otter++~\cite{ahmed2026heterogeneous}, Issue2Test~\cite{nashid2026issue2test}, and AssertFlip~\cite{khatib2026assertflip} in issue resolution.
We run the \doc artifact with \minimax-M3 and \gpt-5.6-Luna, extract localization from its diagnosis records (\doc-L), and, since \doc supports only 107 Python bug-fixing instances in \SweBP, we evaluate all approaches on this subset for a consistent comparison.

Besides \emph{localization outputs before patching} (\name-L, \cc-L, and \doc-L), we extract \emph{localization from each approach's final patch} \name-P, \miniSA-P, \cc-P, and \doc-P), reflecting the influence of downstream repair and validation. Since \miniSA exposes no separate localization output, only \miniSA-P is reported.
% This comparison therefore measures agreement with gold-patch locations using each system's available output: results from \miniSA additionally reflect downstream repair and validation decisions.
% Since \doc only supports $107$ bug-fixing Python instances, for both \name and selected baselines, we obtain their localization results only on the \num{266} Python instances and compute both file-level and method-level localization recall, precision, and F1 score against the ground truth edit sites.

\begin{figure}[t]
    \centering
    \includegraphics[width=0.90\textwidth]{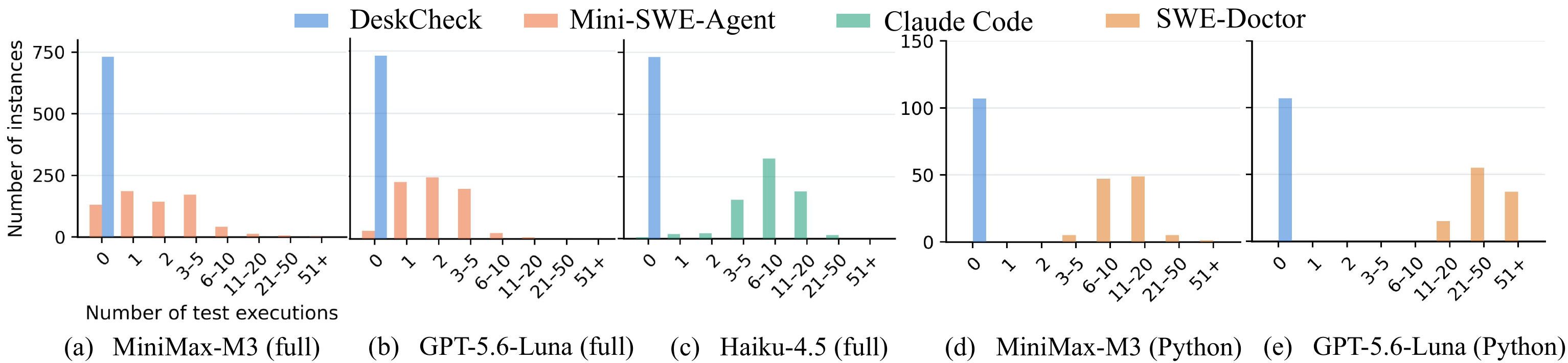}
    \vspace{-10pt}
    \caption{Number of executed reproduction tests before patching.}
    \label{fig:reproduction-test-count}
    \vspace{-20pt}
\end{figure}

As shown in \Cref{tab:localization}, \name achieves the highest resolution rate with every LLM, and \name-L consistently achieves the highest method-level recall, indicating that \textbf{pure reasoning-based localization identifies more potential buggy sites}; this advantage persists after repair, audit, and validation (\name-P).
At the file level, \name-L and \name-P also achieve the highest or near-highest recall (within $0.02$ of the best with \gpt-5.6-Luna), at the cost of lower precision than some baselines, which is acceptable since missing a buggy site harms downstream stages more than proposing extra candidates.
\textbf{\name-L also consistently outperforms \doc-L in recall}, suggesting that guided code reasoning localizes bugs more broadly than \doc's reproduction tests.
Notably, \doc-P recalls far more sites than \doc-L, and \doc resolves more instances than \miniSA, suggesting that its patch generator benefits more from runtime failure symptoms than from localization.

\begin{wraptable}{r}{0.5\columnwidth}
\vspace{-10pt}
\centering
\footnotesize
\caption{Results of the chi-square test on the correlation between the code reasoning-triggered edits at each stage (patch refinement, code audit, validation) and the rescue of the previously failed cases.}
\label{tab:chi-test}
\setlength{\tabcolsep}{3pt}
\resizebox{\linewidth}{!}{%
\begin{tabular}{l|cc|cc|cc}
\hline
\multirow{2}{*}{\textbf{Model}} & \multicolumn{2}{c|}{\textbf{Refine}} & \multicolumn{2}{c|}{\textbf{Audit}} & \multicolumn{2}{c}{\textbf{Validation}} \\ \cline{2-7}
              & $\chi^2$ & p-value & $\chi^2$ & p-value & $\chi^2$ & p-value \\ \hline
\minimax-M3   & 11.92    & 0.001 & 4.84     & 0.028   & 10.10     & 0.002   \\
\gpt-5.6-Luna & 4.68     & 0.030   & 9.89     & 0.002   & 3.85     & 0.047   \\
\haiku-4.5    & 11.27     & 0.001  & 11.90   & 0.001  & 7.73     & 0.005 \\ \hline
\end{tabular}%
}
\vspace{-10pt}
\end{wraptable}

To quantify each baseline's reliance on reproduction tests, we count the actions that execute reproduction tests before patching, using the process-centric analysis tool of \cite{liu2026online} on the agent trajectories and \doc's reports (\Cref{fig:reproduction-test-count}).
Counting repeated executions, \miniSA executes $1.84$ and $2.27$ reproduction tests per instance on average with \minimax-M3 and \gpt-5.6-Luna, \cc $8.51$, and \doc the most: \num{12.6} and \num{43.4}, respectively.
In contrast, \name replaces pre-patching test execution with reasoning-based simulation, avoiding both the cost of running tests and the difficulty of setting up executable environments.

% \begin{rqsummary}
% Finding 2: \name replaces the execution of pre-patching reproduction tests with a pure reasoning-based execution path simulation, which even achieves higher recall rate on both method-level and file-level when compared with the ground truth edit sites
% \end{rqsummary}

\begin{wrapfigure}{r}{0.50\textwidth}
    \vspace{-15pt}
    \centering
    \includegraphics[width=0.50\textwidth]{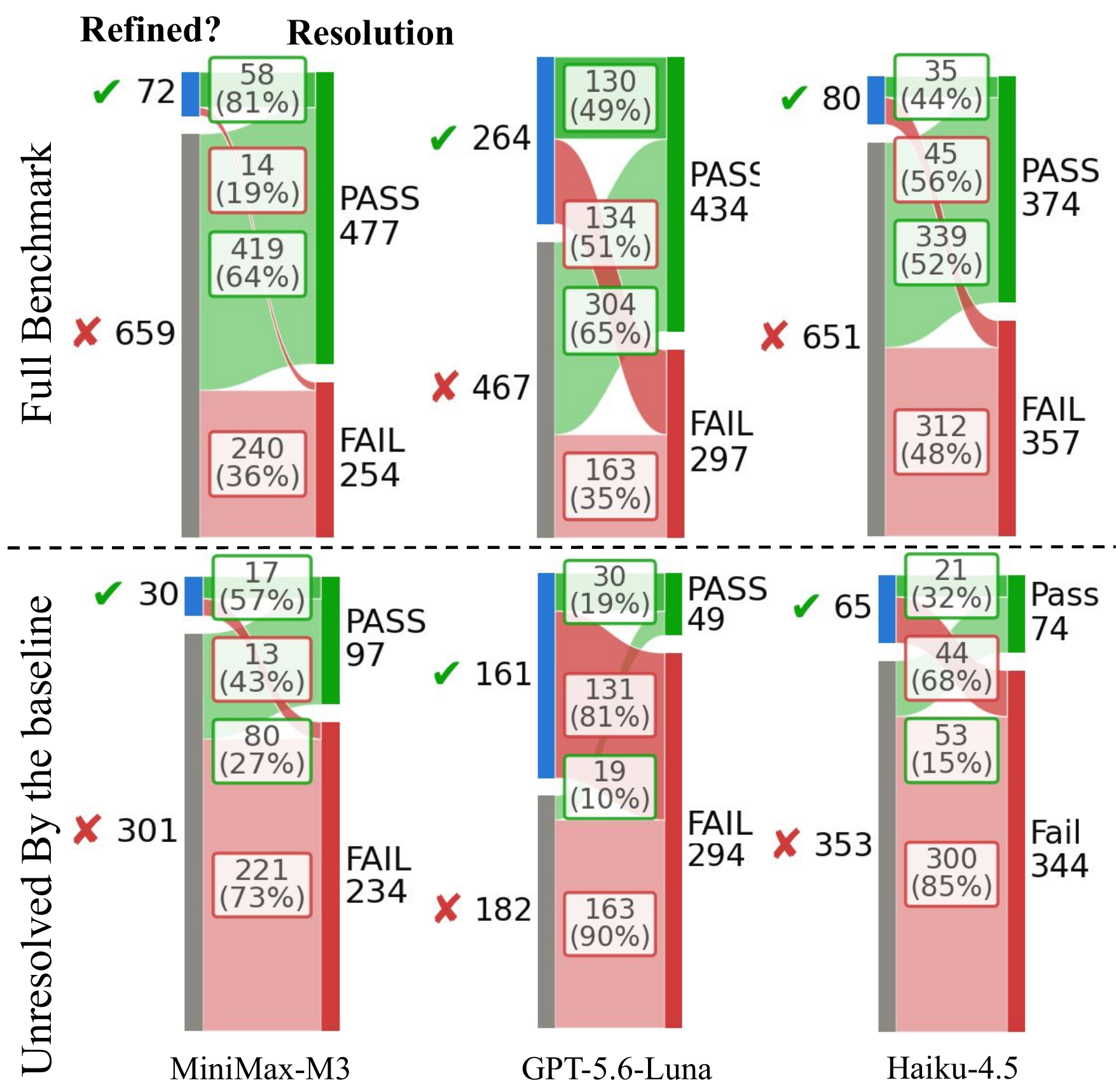}
    \caption{Effectiveness of Repair \& Refine.}
    \label{fig:refine-effectiveness}
    \vspace{-15pt}
\end{wrapfigure}

\subsubsection{Code Reasoning in Repair \& Refine}
As shown in \Cref{fig:refine-effectiveness}, \name refines the initial patch after simulation on $72$, \num{264}, and $80$ instances with \minimax-M3, \gpt-5.6-Luna, and \haiku-4.5, of which $58$ ($80.6\%$), \num{130} ($49.2\%$), and $35$ ($43.8\%$) are resolved, respectively.
Among the \num{331}, \num{343}, and \num{418} instances that the baselines fail to resolve, refinement is triggered on $30$, \num{161}, and $65$, rescuing $17$, $30$, and $21$ instances, a higher rescue rate than without refinement (\minimax-M3:$57\%$ vs.\ $27\%$, \gpt-5.6-Luna: $19\%$ vs.\ $10\%$, and \haiku-4.5: $32\%$ vs.\ $15\%$).
We run a chi-square test whose null hypothesis is that patch refinement and the rescue of previously failed instances are independent.
The test (\Cref{tab:chi-test}) rejects it and confirms a \textbf{statistically significant correlation between patch refinement and the rescue of previously failed instances} for all three models ($p<0.05$).

% Keep this figure off the current page so it lands at the top of the next one.
\suppressfloats[t]
\begin{figure}[t]
\centering
    \includegraphics[width=0.85\textwidth]{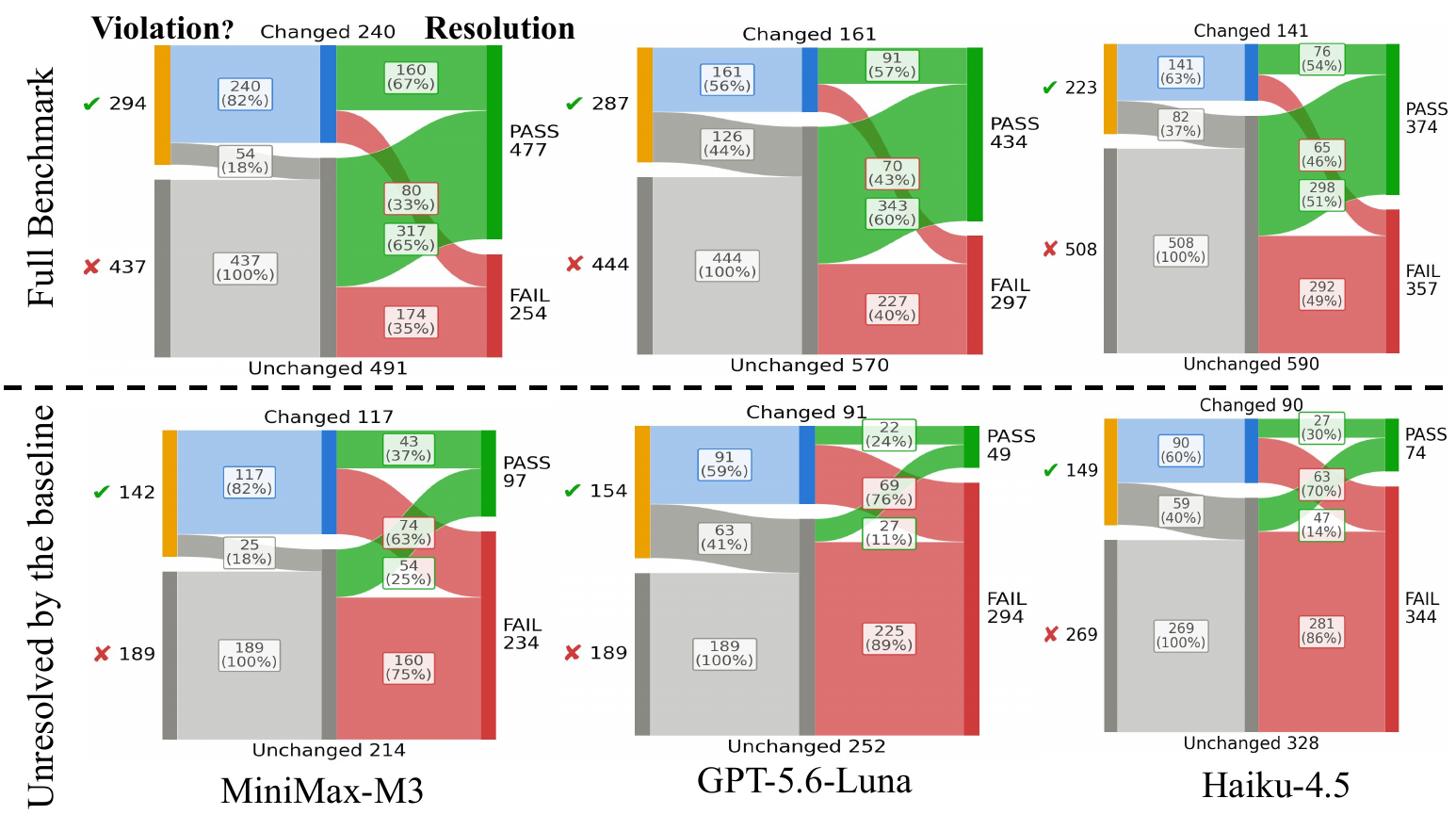}
    \vspace{-10pt}
    \caption{Effectiveness of the audit stage driven by the patch simulation.}
    \label{fig:audit-effectiveness}
    \vspace{-20pt}
\end{figure}

\subsubsection{Code Reasoning in Audit}
\label{sec:effectiveness-audit}
\vspace{-5pt}
As traced in \Cref{fig:audit-effectiveness}, after reasoning about whether the patch satisfies each requirement in the issue, \name flags violations in \num{294}, \num{287}, and \num{223} instances with \minimax-M3, \gpt-5.6-Luna, and \haiku-4.5; the lightweight repair changes \num{240}, \num{161}, and \num{141} of these patches, of which \num{160}, $91$, and $76$ are resolved, respectively.
Of the \num{262} violations that the lightweight repair dismisses as false alarms, \num{126} are still resolved, justifying its right to refute them.
On the instances that \miniSA and \cc fail to resolve, patches changed in the audit stage are rescued more often than untouched ones (\minimax-M3: $37\%$ vs.\ $25\%$, \gpt-5.6-Luna: $24\%$ vs.\ $11\%$, \haiku-4.5: $30\%$ vs.\ $14\%$).
A chi-square test (\Cref{tab:chi-test}) confirms a \textbf{statistically significant correlation between the code reasoning-driven audit and the rescue of previously failed instances} for all three models.

\begin{wrapfigure}{r}{0.45\textwidth}
    \vspace{-20pt}
    \centering
    \includegraphics[width=0.45\textwidth]{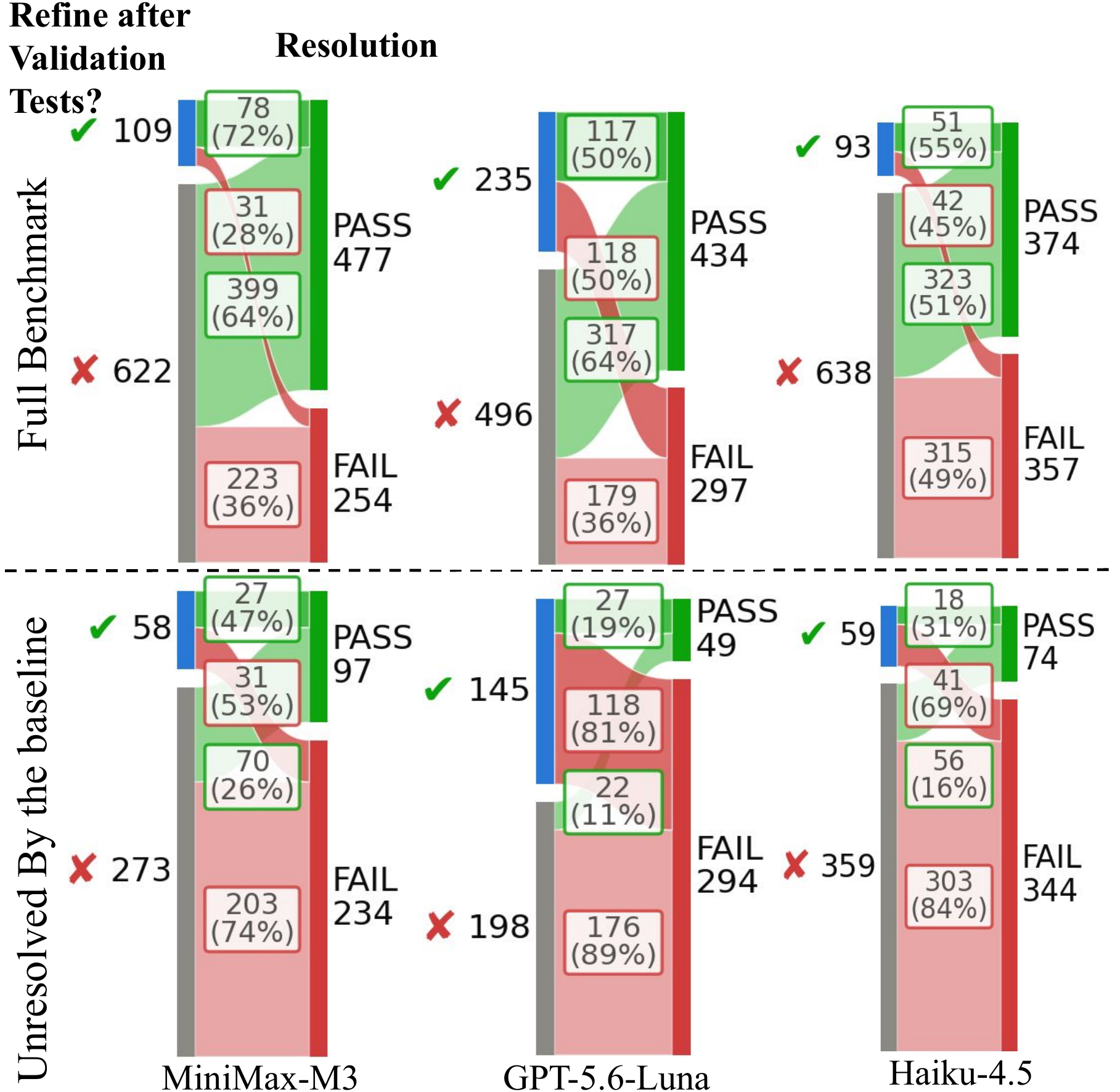}
    \vspace{-20pt}
    \caption{Effectiveness of the code reasoning-augmented validation.}
    \label{fig:validation-effectiveness}
    \vspace{-15pt}
\end{wrapfigure}

\subsubsection{Code Reasoning in Validation}
As shown in \Cref{fig:validation-effectiveness}, \name revises the patch after generated tests fail on \num{109}, \num{235}, and \num{93} instances with \minimax-M3, \gpt-5.6-Luna, and \haiku-4.5, of which $78$, $117$, and $51$ are resolved, respectively.
On the instances that the baselines fail to resolve, \textbf{patches edited in the validation stage are rescued more often than unedited ones} (\minimax-M3: $47\%$ vs.\ $26\%$, \gpt-5.6-Luna: $19\%$ vs.\ $11\%$, \haiku-4.5: $31\%$ vs.\ $16\%$).
A chi-square test (\Cref{tab:chi-test}) similarly rejects the independence between validation-stage edits and the rescue of previously failed instances for all models ($p<0.05$).
These revisions are driven by tests built from reasoned input-expected output pairs, which cover more diverse scenarios and edge cases that may reveal the defects in the original patch.

\section{Qualitative Analysis}
\label{sec:qualitative}

\begin{figure}[t]
    \centering
    \includegraphics[width=0.99\textwidth]{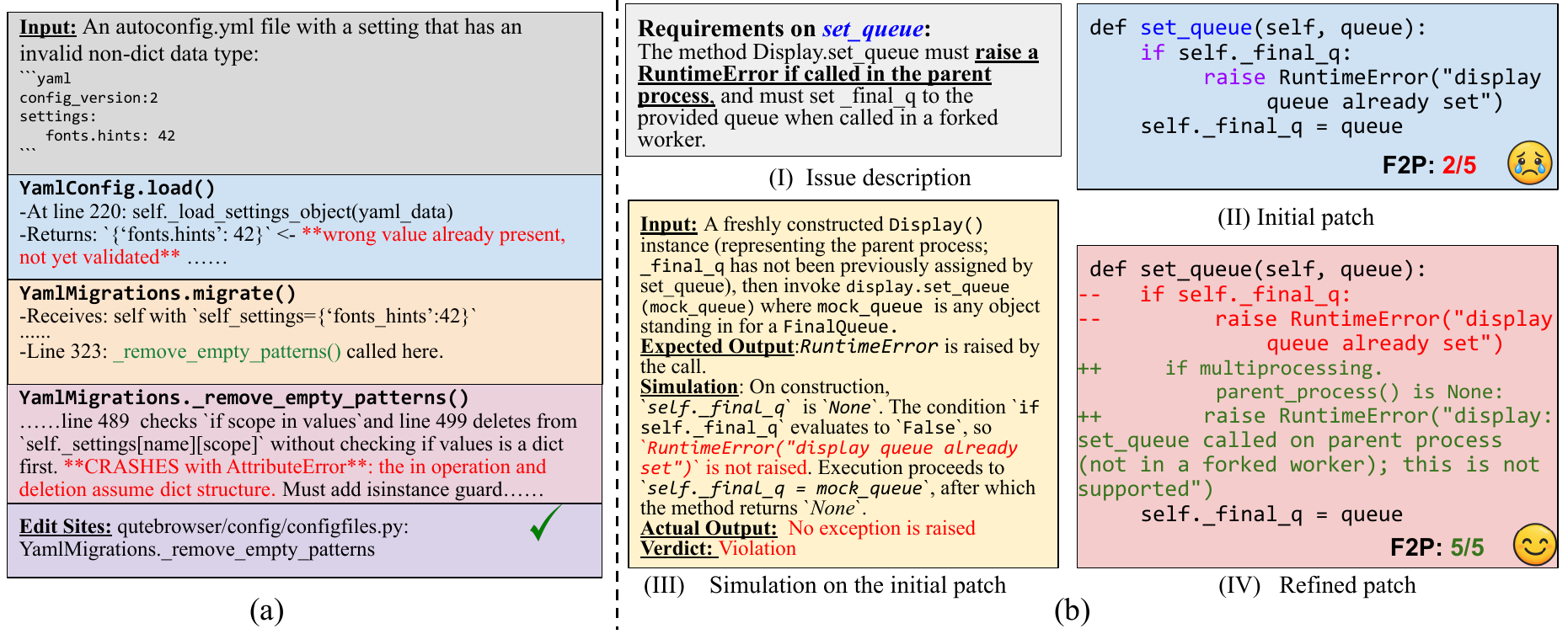}
    \vspace{-10pt}
    \caption{(a) An example of \name(\haiku-4.5) locating the root cause of the bug; (b) An example of \name(\minimax-M3) correctly refines the initial patch after code reasoning.}
    \label{fig:quanlitative-example-localization-refine}
    \vspace{-15pt}
\end{figure}

\subsection{Execution path simulation improves localization}
Referring back to \Cref{fig:reproduction-surface}, 
\name selects \texttt{YamlConfig.load()} $\rightarrow$ \texttt{YamlMigrations.migrate()} $\rightarrow$ \texttt{YamlMigrations.\_remove\_empty\_patterns()} 
into $S_B$ in \Cref{fig:quanlitative-example-localization-refine}a, which is neglected by the baseline.
\Cref{fig:quanlitative-example-localization-refine}a also shows how \name reasons about this path: it creates an input file with concrete keys and values, correctly infers how the invalid \texttt{settings} is constructed in \texttt{YamlConfig.load()}, and traces the subsequent execution until \texttt{\_remove\_empty\_patterns()}, then recognizes that the \texttt{in} and \texttt{del} operations would crash on the invalid \texttt{settings}.
\name therefore adds \texttt{\_remove\_empty\_patterns} to the edit sites, a function called by the hidden tests but never mentioned in the issue description and missed by the baseline \cc.

\subsection{Initial patch simulation saves near-hits}
Referring back to \Cref{fig:near-miss}, the initial patch (\Cref{fig:quanlitative-example-localization-refine}b-II) raises the \texttt{RuntimeError} when the display queue is already set, rather than when it is called in a parent process as the issue description requires (\Cref{fig:quanlitative-example-localization-refine}b-I).
\Cref{fig:quanlitative-example-localization-refine}b-III shows how \name simulates this patch: \minimax-M3 infers an input in which \texttt{set\_queue} is called from a parent process, a condition that is hard to reproduce with real tests since it requires setting up and controlling multiple processes. It then notices that the \texttt{RuntimeError} would not be raised as expected and flags a violation in the initial patch.
\name therefore routes back to the lightweight repair and adjusts the condition in the error handling, so the refined patch passes all the hidden tests (\Cref{fig:quanlitative-example-localization-refine}b-IV), whereas the baseline \miniSA fails to set up such a test after 14 steps and submits the near-hit untested.

\begin{wrapfigure}{r}{0.50\textwidth}
    \vspace{-10pt}
    \centering
    \includegraphics[width=0.50\textwidth]{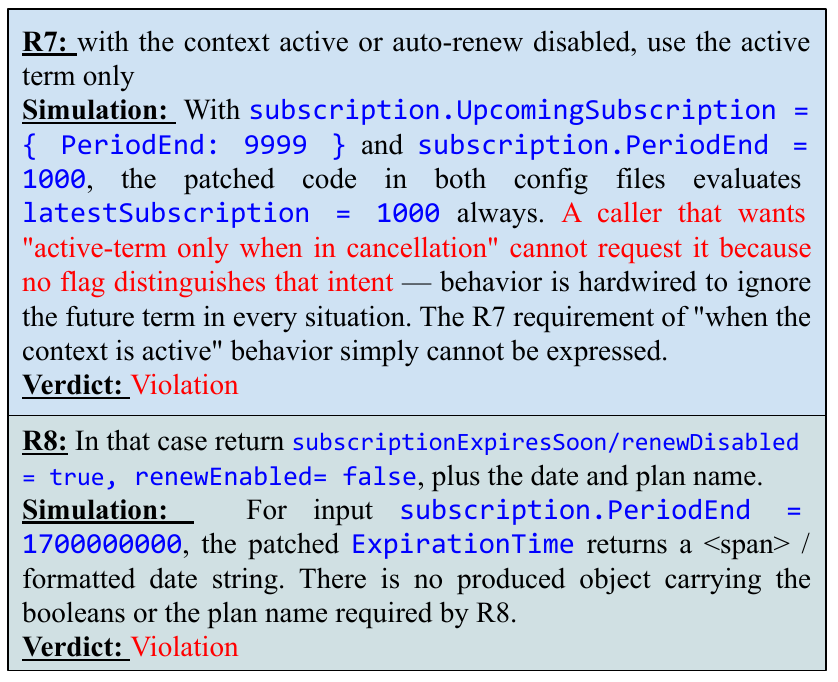}
    \vspace{-25pt}
    \caption{An example of \name(\gpt-5.6-Luna) auditing against the requirements to complete the patch.}
    \label{fig:quanlitative-example-audit}
    \vspace{-20pt}
\end{wrapfigure}

\subsection{Auditing against the requirements completes the patch}
\Cref{fig:quanlitative-example-audit} shows how \name performs point-by-point code reasoning against the requirements to rescue the incomplete patch.
Referring back to \Cref{fig:omitting-requirements}, the patch before the code auditing phase only fulfills \num{6} out of the \num{8} extracted requirements, leaving one file totally unchanged.
For both R7 and R8, in the code auditing sub-agent, \name simulates the execution of the patch by creating a concrete input/scenario that satisfies the conditions inferred from the requirement, judging whether the current patch behaves as expected.
It finally raises two violations corresponding to R7 and R8, guiding the agent to conduct more repair on the patch.
Finally, the post-audit refinement flips the fail-to-pass rate from 0/1 to 1/1.

\begin{wraptable}{r}{0.65\textwidth}
\vspace{-12pt}
\centering
\scriptsize
\setlength{\tabcolsep}{3pt}
\renewcommand{\arraystretch}{0.95}
\caption{Oracle comparison between \SweBP hidden tests and validation tests generated by \name(\minimax-M3).}
\label{tab:qualitative-oracle}
\vspace{-10pt}
% Light green: oracle present; light red: oracle missing (NA).
\definecolor{oraclegreen}{RGB}{226,240,217}
\definecolor{oraclered}{RGB}{248,215,215}
\begin{tabular}{|c|l|l|}
\hline
\textbf{id} & \textbf{Hidden test oracles} & \textbf{\name oracles} \\ \hline
t1 & \cellcolor{oraclegreen}("DVD", NONBOOK) $\to$ True & \cellcolor{oraclegreen}("DVD", NONBOOK) $\to$ True \\ \hline
t2 & \cellcolor{oraclegreen}("dvd", NONBOOK) $\to$ True & \cellcolor{oraclegreen}("dvd", NONBOOK) $\to$ True \\ \hline
t3 & \cellcolor{oraclegreen}("audio cassette", NONBOOK) $\to$ True & \cellcolor{oraclegreen}("Audio CD", NONBOOK) $\to$ True \\ \hline
t4 & \cellcolor{oraclegreen}("paperback", NONBOOK) $\to$ False &
  \cellcolor{oraclegreen}\begin{tabular}[c]{@{}l@{}}("paperback", NONBOOK) $\to$ False\\ ("Hardcover", NONBOOK) $\to$ False\end{tabular} \\ \hline
t5 & \cellcolor{oraclered}\textcolor{red}{NA} &
  \cellcolor{oraclegreen}\begin{tabular}[c]{@{}l@{}}(None, NONBOOK) $\to$ False\\ ('', NONBOOK) $\to$ False\end{tabular} \\ \hline
\end{tabular}
\vspace{-12pt}
\end{wraptable}

\subsection{Reasoning about inputs and expected outputs improves validation test quality}
\label{sec:qualitative-validation}
Referring back to \Cref{fig:insufficient-validation}, the validation tests of \miniSA for \texttt{is\_nonbook} in openlibrary-3f7db6bb never consider case-insensitive (t2) or multi-word (t3) inputs.
\Cref{tab:qualitative-oracle} compares the oracles in the validation tests of \name with those in the hidden tests: by reasoning about input-expected output pairs before generating tests, \name independently covers a case-insensitive input that matches hidden test t2 and a multi-word input (t3), and further adds empty and \texttt{None} inputs (t5) that neither the baseline nor the hidden tests consider.
\name therefore catches that the initial patch fails t2 in the first run and refines it to pass, whereas \miniSA submits a patch that fails the hidden tests.

\vspace{-5pt}
\section{Ablation Study}
\label{sec:ablation}
\begin{wraptable}{r}{0.44\textwidth}
\vspace{-12pt}
\centering
\scriptsize
\setlength{\tabcolsep}{2pt}
\renewcommand{\arraystretch}{0.95}
\caption{Ablation results. Loc./Rep./Aud./Val. indicate whether code reasoning is enabled in the Localization, Repair, Audit, and Validation stages; Res. is the resolution rate and Rec. the method-level localization recall.}
\label{tab:ablation-settings}
\vspace{-8pt}
\begin{tabular}{@{}lcccccc@{}}
\toprule
\textbf{Setting} & \textbf{Loc.} & \textbf{Rep.} & \textbf{Aud.} & \textbf{Val.} & \textbf{Res.} & \textbf{Rec.} \\
\midrule
$V_0$ & \xmark & \xmark & \xmark & \xmark & 53\% & 0.59 \\
$V_1$ & \cmark & \xmark & \xmark & \xmark & 56\% & 0.65 \\
$V_2$ & \cmark & \cmark & \xmark & \xmark & 58\% & 0.66 \\
$V_3$ & \cmark & \cmark & \cmark & \xmark & 61\% & \textbf{0.67} \\
\name & \cmark & \cmark & \cmark & \cmark & \textbf{68\%} & \textbf{0.67} \\
\midrule
\miniSA & -- & -- & -- & -- & 53\% & 0.57 \\
\bottomrule
\end{tabular}
\vspace{-14pt}
\end{wraptable}

To isolate the contribution of code reasoning, we disable it stage by stage: $V_0$ removes it from all sub-agents; $V_1$ restores it in localization; $V_2$ further restores it in repair, triggering refinement of the initial patch; and $V_3$ further restores it in the audit, where each requirement is checked by simulating the patch on specific inputs.
Each variant disables \emph{only the explicit code reasoning step}, retaining all other operations (e.g., patch refinement, requirement extraction, and test generation and execution).
As running every variant on all $731$ \SweBP instances is prohibitively expensive, we evaluate on a 100-instance subset designed to reflect the benchmark's composition.
We stratify instances by language and difficulty (\S\ref{sec:rq-empirical-factors}), and within each language-difficulty group, balance the sample across repositories and randomly select instances from each.

With \minimax-M3, \Cref{tab:ablation-settings} reports the resolution rate and method-level localization recall of the variants, \name, and \miniSA: \textbf{the resolution rate increases monotonically from $\mathbf{53\%}$ ($V_0$) to $\mathbf{56\%}$ ($V_1$), $\mathbf{58\%}$ ($V_2$), and $\mathbf{61\%}$ ($V_3$), reaching $\mathbf{68\%}$ with the full system}.
Every restored stage contributes to the resolution rate, and restoring code reasoning in validation brings the largest gain ($61\%$ to $68\%$), indicating that reasoning about inputs and expected outputs before generating tests is particularly effective.
$V_0$ performs similarly to \miniSA, indicating that \name's improvement comes from code reasoning across multiple phases rather than from isolating contexts into sub-agents; meanwhile, method-level recall improves from $0.59$ to $0.65$ once code reasoning is restored in localization ($V_1$) and reaches $0.67$ thereafter, suggesting that simulating execution paths identifies more bug-relevant locations.

\vspace{-5pt}
\section{Related Work}
\textbf{Code Reasoning Evaluations.}
Existing benchmarks evaluate the code reasoning~\cite{abdollahi2025demystifying, majgaonkar2025understanding} of LLMs via input/output prediction~\cite{gu2024cruxeval}, specification and dynamic-semantics reasoning~\cite{liu2026codemind}, and intermediate program states~\cite{chen2024runtime, liu2025assessing, beger2025coconut}, while another line of work analyzes agent trajectories through thought--action--result sequences~\cite{bouzenia2025understanding}, action graphs (Graphectory~\cite{liu2026process,liu2026plan}), or combined formal and LLM-based reviews (AgentLens~\cite{podivilov2026agentlens}).
However, the former targets standalone programs, and the latter focuses on actions and tool usage rather than reasoning; \textbf{we instead study the role of code reasoning in the problem-solving process of programming agents at scale.}

\noindent\textbf{Reproduction Tests for Bug Localization.}
% \alireza{the only reference I have related to this subject is this one https://arxiv.org/abs/2606.26978 I believe it's not closely related to our work}
% I added more related work on reproduction test generation and revised the paragraph accordingly, including removing AutoCodeRover.
%(1) generation --> (2) refinement --> (3) diagnosis/localization
%(1) generation
Reproduction tests translate issue descriptions into executable evidence of reported failures. AEGIS~\cite{wang2024aegis} and Otter~\cite{ahmed2025otter} generate them directly from issue reports, e-Otter++~\cite{ahmed2026heterogeneous} adds heterogeneous prompting and execution feedback for test generation and selection, Issue2Test~\cite{nashid2026issue2test} refines candidate tests with compilation and runtime feedback until their failures match the issue, and AssertFlip~\cite{khatib2026assertflip} inverts the assertions of passing tests on buggy behavior.
%(2) refinement
%(3)diagnosis/localization
Programming agents further use such tests as execution evidence: MarsCode Agent~\cite{liu2024marscode} uses exception stacks and outputs of reproduction scripts, and SWE-Doctor~\cite{guo2026swedoctor} turns tests targeting different behavioral requirements into runtime diagnoses for localization and repair.
% Prior agents use reproduction tests as execution evidence for localization: MarsCode Agent~\cite{liu2024marscode} debugs with exception stacks and outputs of generated reproduction scripts, AEGIS~\cite{wang2024aegis} refines such scripts via structured context extraction and iterative feedback, and SWE-Doctor~\cite{guo2026swedoctor} generates tests targeting different behavioral requirements and combines their runtime diagnoses with localization hints.
% \jiyoung{
% e-Otter++~\cite{ahmed2026heterogeneous} improves reproduction test generation and selection through heterogeneous prompting and execution feedback, Isse2Test~\cite{nashid2026issue2test} iteratively refines candidate tests using compilation and runtime feedback until their failures align with the reported issues. AssertFlip~\cite{khatib2026assertflip} generates passing tests for buggy behavior and inverts their assertions to produce bug-reproducing tests. 
% }
% AutoCodeRover~\cite{zhang2024autocoderover} additionally applies spectrum-based fault localization when a test suite is available.
However, given ambiguous or complex issues, reproduction tests may capture only surface-level symptoms; \textbf{\name instead traces execution paths via code reasoning to reveal the root cause.}
%  \reyhan{a huge body of work on reproduction test generation from issue descriptions is missing here. Remove autocoderover}

\noindent\textbf{Patch Refinement.}
%  \alireza{I see this one https://arxiv.org/abs/2604.03253 it might be useful.}
A coding agent's first patch is often close to, but not exactly, the correct fix~\cite{liu2024automated}, motivating refinement: PatchPilot~\cite{pmlr-v267-li25cf} re-enters generation with failing reproduction or regression tests, REFINE~\cite{pabba2025refine} enriches repository context and reranks candidate patches, and Luo et al.~\cite{luo2026recap} refine patches toward conciseness.
These signals come from \emph{outside} the patch and are bounded by availability and quality: reproduction tests may be missing or unrunnable in unconfigured environments~\cite{xiang2026evaluating}, and over-narrow self-generated tests can give a false sense of correctness~\cite{wang2026solved}.
\textbf{In contrast, \name derives the signal from the patch itself by simulating it on inputs inferred from the issue, which works even without a runnable test.}

\section{Threats to Validity}

\noindent\textbf{External Validity.}
Our conclusions on code reasoning are drawn from \num{7955} trajectories produced by six scaffolds and three model families on the SWE-bench family, so the results may not transfer to other scaffolds, more recent LLMs.
To mitigate this, we deliberately select scaffolds spanning diverse architectures, ranging from plain agents to multi-agent, memory-augmented, and orchestrator--sub-agent designs, and we collect results on \SweBP, which covers Python, JavaScript, Go, and TypeScript, rather than the Python-dominating \SweBV and \SweBL (\S\ref{sec:empirical-study}).

\noindent\textbf{Internal Validity.}
LLM-based systems are inherently non-deterministic, so is \name.
To further verify the contribution of code reasoning instead of non-determinism,
we (1) use an exact McNemar test to verify the significance of improvements, and (2) rerun each baseline twice on the instances that only \name resolves, confirming that $83.18\%$ of them are not explained by run-to-run non-determinism (\S\ref{sec:non-determinism}).

\noindent\textbf{Construct Validity.}
The hidden tests of \SweBP may reject correct patches.
Therefore, we manually verify the false negatives of \name and \miniSA (\S\ref{sec:human-verification}).
\vspace{-10pt}
\section{Conclusion \& Future Work}
In this work, we derive a systematic categorization of the reasoning behaviors exhibited by programming agents.
Then we perform a large-scale empirical study with \num{7955} trajectories, indicating that code reasoning is naturally integrated into the problem-solving process of programming agents.
Then we propose \name, a five-stage agent framework that leverages different levels of code reasoning to improve LLMs' performance in localization, repair, and validation.
Extensive evaluation on \SweBP shows that \name brings up to a $10.53\%$ performance gain compared with \miniSA and \cc, with only a marginal increase in token consumption and cost.
In our future work, we plan to (1) expand the idea of \name to tasks beyond bug-fixing, and (2) certify the faithfulness of  code reasoning, specifically the correctness of execution simulation.

%%
%% Required by the FSE 2027 open science policy: a statement on the
%% availability of a replication package, placed after the Conclusion.
%% It does not count toward the page limit and must remain anonymous.
\section*{Data Availability}
The artifacts of \name are publicly available at \cite{simagent2026artifact}.
%%
%% No acknowledgements at submission time (double-anonymous review).
%% Add an \begin{acks}...\end{acks} block for the camera-ready version,
%% including any required disclosure of generative AI use.

\bibliographystyle{ACM-Reference-Format}
\bibliography{reference}

\end{document}